\documentclass[11pt,a4paper,twoside]{article}
\usepackage[top=1in, bottom=1in, left=1in, right=1in]{geometry}
\usepackage[title,toc,page]{appendix} %Appendixstyle
\usepackage{titlesec}
\usepackage[utf8]{inputenc} % Required for inputting international characters
\usepackage[T1]{fontenc} % Output font encoding for international characters
\usepackage{fancyhdr}
\usepackage{authblk}
\usepackage[colorlinks,citecolor=blue,linkcolor=blue]{hyperref}

\usepackage{mathtools}
\usepackage{amsthm}
\usepackage{mathrsfs}
\usepackage{amsmath}
\usepackage{amsfonts}
\usepackage{amssymb}
\usepackage{yfonts}
\usepackage{BOONDOX-calo}
\usepackage[shortlabels]{enumitem}
\usepackage{braket}
\newtheorem{remark}{Remark}[section]
\newtheorem{Def}{Definition}[section]

\newtheorem{theorem}{Theorem}[section]

\newtheorem{Prop}{Proposition}[section]

\newtheorem{Lemma}{Lemma}[section]

\usepackage{graphicx}
\usepackage{tikz-cd}
\usepackage{color}
\usetikzlibrary{arrows.meta}
\usetikzlibrary{positioning}
\usepackage{booktabs}
\usepackage{subcaption}

\usepackage[autostyle=true,english=american]{csquotes} 
\usepackage[backend=bibtex,natbib=false,style=numeric]{biblatex} % Use the bibtex backend with the authoryear citation style (which resembles APA)
\renewbibmacro{in:}{}
\bibliography{references_used.bib} % The filename of the bibliography

\newcommand{\reeb}{\mathcal{R}} %Reeb
\newcommand{\cH}{\mathcal{H}} %Hamiltonian Function
\newcommand{\ch}{\mathcal{h}}

\newcommand{\Lie}{\mathcal{L}}

\allowdisplaybreaks

\mathtoolsset{showonlyrefs}
\title{Koopman von Neumann formalism meets contact Hamiltonian systems}

\author[1]{Federico Zadra}
\affil[1]{Department of Mathematics, University of Antwerp, Antwerpen (BE)}

\date{}

\begin{document}
\maketitle
\begin{abstract}
    The present work develops a Koopman--von Neumann (KvN) formulation of contact Hamiltonian dynamics on the Hilbert space defined by the contact volume. The resulting generator includes the divergence correction required for unitary evolution. We establish symmetry and, for complete contact Hamiltonian vector fields, essential self-adjointness and a strongly continuous unitary evolution; we also describe covariance under contactomorphisms. For completely integrable contact systems with complete flows, it is possible to recover an invariant direct-integral decomposition and strongly commuting KvN generators. We further relate contact KvN dynamics to the KvN
    formulation of its symplectization.
\end{abstract}

\section{Introduction}
Classical and quantum mechanics in their traditional and most widely used formulation~\cite{Arnold_1989, libermann_symplectic_1987, hall2013quantum} rely on two completely different mathematical structures: the first, is based on the Hamiltonian formulation of the dynamics over a symplectic manifold where Hamiltonian vector fields generate the dynamics, while the quantum dynamics is described in terms of Hilbert spaces, and the generators are self-adjoint operators. At the beginning of the 1930s Koopman~\cite{Koopman1931} and von Neumann~\cite{v_Neumann_1932,Neumann_1932} introduced almost at the same time a formulation of classical mechanics which uses a Hilbert space of arena instead of the traditional symplectic setting. For a system whose phase space is a symplectic manifold $(S,\omega)$, its main features can be summarized as follows:
\begin{itemize}
    \item Classical states are represented by normalized elements $\psi$ of the Hilbert space $\mathscr H=L^2(S,\Omega_S)$, where $\Omega_S$ is the symplectic volume form.
    \item In canonical coordinates, the position and momentum operators $\hat Q$ and $\hat P$ act by multiplication and therefore commute.
    \item The squared modulus $\rho=|\psi|^2$ represents the classical probability density.
    \item The evolution is governed by the KvN equation, $\imath\partial_t\psi=\hat L_h\psi$, where $\hat L_h=-\imath Y_h$ is the Liouville operator. The corresponding probability density satisfies the classical Liouville equation.
    \item Classical observables are represented by self-adjoint multiplication operators associated with real-valued functions on phase space.
\end{itemize}

Although this formulation replaces finite-dimensional classical dynamics with evolution on an infinite-dimensional Hilbert space, it provides a common mathematical language for classical and quantum mechanics. In particular, KvN methods have been employed to investigate classical-quantum hybrid dynamics and the coupling between classical and quantum systems~\cite{GayBalmazTronci2022}. The KvN formalism can be employed to introduce the classical path-integral formulation, which provides a functional-integral representation of classical Hamiltonian dynamics~\cite{gozzibook}.

The KvN approach is not limited to symplectic Hamiltonian systems and has already been generalized to nonlinear and non-Hamiltonian dynamics~\cite{Joseph2020}. Nevertheless, contact Hamiltonian mechanics provides a particularly interesting geometric framework for describing dissipative systems~\cite{Bravetti2017}. Unlike symplectic Hamiltonian vector fields, contact Hamiltonian vector fields are not generally divergence free with respect to the natural contact volume form. Although invariant measures may exist under suitable conditions~\cite{Bravetti_2015,Bravetti_2020}, the contact volume is not necessarily preserved by the dynamics. Therefore, the standard symplectic KvN generator cannot be directly extended simply by replacing the underlying phase space. Instead, the evolution operator must account for the divergence of the contact Hamiltonian vector field to ensure unitarity. In this work, we investigate this construction from the perspective of contact geometry, characterizing its operator-theoretic properties and its relationship with contact Hamiltonian dynamics.

The paper is organized as follows. Section~\ref{sec:background} recalls the necessary symplectic and contact geometry. Section~\ref{sec:contacKoopman} develops the contact Koopman-von Neumann construction and its spectral and entropic properties. Section~\ref{sec:KvNexamples} examines the Reeb flow and damped examples; the appendix reviews the symplectization convention used in the comparison.

%%%%%%%%%%%%%%%%%%%%%%%%%%%%%%%%%%%%%%%%%%%%%%%%%%%%%%%%%%%%%%%%%%%%%%%%%%%%%%%%%%%%%%%%%%
\section{Background}
\label{sec:background}
%%%%%%%%%%%%%%%%%%%%%%%%%%%%%%%%%%%%%%%%%%%%%%%%%%%%%%%%%%%%%%%%%%%%%%%%%%%%%%%%%%%%%%%%%%
In this section, we introduce the elements necessary to deal with the contact formulation of KvN mechanics, we start reviewing the basics of classical mechanics and the related KvN  approach, and then move to contact geometry and the different aspects of the Hamiltonian dynamics.

%%%%%%%%%%%%%%%%%%%%%%%%%%%%%%%%%%%%%%%%%%%%%%%%%%%%%%%%%%%%%%%%%%%%%%%%%%%%%%%%%%%%%%%%%%
\subsection{Symplectic manifolds and classical mechanics}
\label{sec:introkvn}
The Hamiltonian perspective of classical mechanics consists in identifying a generalized coordinate manifold $E$ and then working on its cotangent bundle $T^*E$. Any cotangent bundle presents a natural (exact) symplectic structure $ - d \lambda$, where $\lambda$ is the Liouville form~\cite{Arnold_1989}. The dynamics of these systems is usually induced by a function, called Hamiltonian
\begin{equation}
    h : T^*E \to \mathbb{R},
\end{equation}
which induces a vector field $Y_h$ over $T^*E$ that satisfies:
\begin{equation}
    -(d \lambda)(Y_h, \cdot) = d h.
\end{equation}

This construction can be generalized to any symplectic manifold, which consists of an even dimensional manifold $S$ endowed by a closed non degenearte $2$-form $\omega$. The Hamiltonian vector field $Y_h$ satisfies then
\begin{equation}
    \label{eqn:sympHVF}
    \omega(Y_h, \cdot) = d h.
\end{equation}

The standard coordinates are characterized by the Darboux theorem~\cite{Arnold_1989}, which states that for each point $x \in S$ there exists a neighbourhood $U$ for which there exists a set of local canonical coordinates $(q_1, \ldots, q_n, p_1, \ldots p_n)$ such that the symplectic form $\omega$ can be written as:
    \begin{equation}
        \omega\lvert_U = dq_i \wedge dp^i,
    \end{equation}
    and the Hamiltonian equations take the well known form:
    \begin{equation}
        \begin{dcases}
            \dot{q}_i = \frac{\partial h}{\partial p^i},\\
            \dot{p}_i = - \frac{\partial h}{\partial q_i}.
        \end{dcases}
    \end{equation}

Furthermore, a symplectic manifold owns a natural volume form induced by the symplectic form: if the manifold $S$ has dimension $2n$ the volume form is given by:
\begin{equation}
    \Omega_S := \omega^{n} = \underbrace{\omega \wedge \ldots \wedge \omega}_{n}.
\end{equation}
Using the definition of Hamiltonian vector field, it is possible to prove the Liouville theorem, and it relies mostly on the divergence free nature of any Hamiltonian vector field with respect to the volume form $\Omega$.
\begin{theorem}[Liouville~\cite{huang1987statistical}]
    \label{thm:Liouville}
    The phase space volume is preserved by the symplectic Hamiltonian vector field.
\end{theorem}

%%%%%%%%%%%%%%%%%%%%%%%%%%%%%%%%%%%%%%%%%%%%%%%%%%%%%%%%%%%%%%%%%%%%%%%%%%%%%%%%%%%%%%%%%%%%%
\subsubsection{Liouville equations and Koopman von Neumann mechanics}
\label{subsec:Liouville}
Taking advantage of the natural volume form $\Omega_S$, it is possible to introduce spaces
\begin{equation}
L^p_\omega(S) = \left\{ \rho:S \to \mathbb{R}\, \bigg\lvert \, \int_S \lvert \rho \rvert^p \, \Omega_S < +\infty \right\}. 
\end{equation}
A subset of $L^1_\omega(S)$ is given by the probability distributions: non-negative functions that are normalized to one by the norm of $L^1_\omega(S)$.

Theorem~\ref{thm:Liouville}, together with the requirement that the integral over time of the norm of a probability distribution does not change, i.e. $\partial_t\, \lvert\lvert \rho \rvert\rvert = 0, $ leads to the Liouville equation:
\begin{equation}
    \label{eqn:sympLiouville}
    \partial_t \rho = - Y_h \rho = \{\rho, h\}_\omega;
\end{equation}
and this is a well known result in classical mechanics and classical statistical mechanics~\cite{huang1987statistical}.

This is the starting point of the Koopman-von Neumann formulation of classical mechanics: the idea is to associate to a mechanical system instead of a symplectic manifold and system of ordinary differential equations of Hamiltonian nature, a Hilbert space and a unitary evolution.
Koopman and von Neumann~\cite{Koopman1931, v_Neumann_1932,Neumann_1932} have independently introduced this formalism in the thirties in order to unify the formalism of classical and quantum mechanics:
\begin{equation}
    L^2_\omega(S) := \left\{\psi: S \to \mathbb{C} \, \bigg\lvert \int_S \psi(x) \overline{\psi(x)}\, \omega^n < +\infty \right\},
\end{equation}
where $\overline{\psi(x)}$ indicated the complex conjugate of $\psi(x)$.
Since we are interested in $\psi \in L^2_\omega(S)$ such that their modulus squared is a probability distribution, i.e.
\begin{equation}
    \lvert \psi (x) \rvert^2 = \rho(x) \in L^1_\omega(S), 
\end{equation}
it is possible to recover the Koopman von Neumann equation, equivalent to the Liouville equation~\eqref{eqn:sympLiouville}, that can be seen as the classical analog of the Schr\"odinger equation:
\begin{equation}
    \imath \partial_t \psi(t,{q_i,p^i}) = \hat{L}_h \psi(t,{q_i,p^i}) = - \imath Y_h (\psi(t,{q_i,p^i})),
\end{equation}
where we define the Liouville operator, $\hat{L}_h : = - \imath Y_h$. For a more detail derivation please refer to~\cite{gozzibook,Koopman1931,v_Neumann_1932}.
\begin{remark}
    The most immediate difference between the KvN equation and the Schrodinger equation is that the quantum usually presents diffusion due to the second order kinetic operator, while the classical evolution is purely a first order partial differential equation, i.e. a transport equation.
\end{remark}

%%%%%%%%%%%%%%%%%%%%%%%%%%%%%%%%%%%%%%%%%%%%%%%%%%%%%%%%%%%%%%%%%%%%%%%%%%%%%%%%%%%%%%%%%%%%
\subsection{Dissipative mechanics and Contact geometry}

Contact geometry is the odd dimensional cousin of the symplectic framework. Beyond dimension, the most immediate difference relies on the characterizing structure which is encoded in a distribution on the tangent space(sub-bundle)~\cite{Geiges_2008}. In this work, the focus will be on co-oriented (or exact) contact manifolds, which distribution can be globally identified by kernel of a one-form.
\begin{Def}
    A contact manifold is a couple $(M,\Delta)$, where $M$ is an $(2n+1)$-dimensional manifold and $\Delta\subset TM$ is a co-rank 1 distribution, such that for any point $p \in M$ there exists a neighborhood $U$ in which $\Delta\lvert U = \ker \eta_U$ and
    \begin{equation}
        \eta_U \wedge d{\eta_U}^{\wedge n} \neq 0.
    \end{equation}
	A co-oriented contact manifold is a contact manifold $(M,\Delta)$, on which there globally exist a $1$-form $\eta$, such that $\ker \eta = \Delta$ and
	\begin{equation}
		\Omega_M = \eta \wedge d \eta^{\wedge n},
	\end{equation}
	represent a volume form for $M$.
\end{Def}
During the rest of this work a contact manifold will be a co-oriented contact manifold.
On a contact manifold $(M,\eta)$, the contact form permits a splitting of the tangent bundle into the direct sum of two parts: the kernel of $\eta$ (the horizontal/contact distribution), and the kernel of $d\eta$, $TM = \ker \eta \oplus \ker d \eta$.
The $1$-dimensional subspace, $\ker d \eta$, can be described through the Reeb vector field, which is the unique vector field~\cite{Geiges_2008} satisfying
\begin{equation}
	i_\reeb \eta = 1, \qquad i_\reeb d\eta = 0.
\end{equation}

Transformations over a contact manifold that preserve the kernel of the contact form, and then rescales the contact form by a non zero function, are called contactomorphism. 

\begin{Def}~\cite{Geiges_2008,Bravetti2017}
    If $(M,\eta)$ and $(N,\alpha)$ are two contact manifold, a transformation $\phi:M \to N$ that satisfies:
    \begin{equation}
        \phi^* \alpha = f_\phi \eta,
    \end{equation}
    where $f_\phi: M \to \mathbb{R}^+$ is called contactomorphism (or contact transformation). If the $f_\phi=1$ we call $\phi$ an strict contactomorphism.
\end{Def}
Since the natural volume form $\Omega_M$ depends directly on the contact form $\eta$, it is possible to recover its transformation law.
\begin{Prop}~\cite{Geiges_2008,Bravetti_2020}
    \label{prop:volumetransf}
 Let $\phi : M \to M$ be a contactomorphism, with a scaling function $f_\phi$, then:
    \begin{equation}
        \phi^* \Omega = f_\phi^{n+1} \Omega.
    \end{equation}
\end{Prop}
\subsubsection{Contact Hamiltonian systems.} 
A big analogy to the symplectic setting is the chance to introduce Hamiltonian dynamics. Let $\cH$ be a real smooth function defined on the manifold, that we will refer as Hamiltonian, the dynamics induced by this Hamiltonian function is encoded in the vector field $X_{\cH}$, called Hamiltonian vector field. The Hamiltonian vector field is defined by the two following relations
\begin{equation}
    \label{eqn:defHVF}
	\begin{dcases}
    i_{X_{\cH}} \eta = - \cH, \\ i_{X_\cH} d\eta = dH - \reeb(\cH) \eta,
    \end{dcases}
\end{equation}
the triple $(M,\eta, \cH)$ is called contact Hamiltonian system.

Exploiting equations~\eqref{eqn:defHVF} it is possible to see that:
\begin{equation}
    \mathcal{L}_{X_\cH} \eta = - \reeb(\cH) \eta,
\end{equation}
and consequently recover the transformation law of the volume form $\Omega_M$.

\begin{theorem}[\cite{Bravetti2017}]
    \label{thm:scalingvolume}
    The flow of a Hamiltonian vector field rescales the volume form, i.e. if $X_\cH$ is a contact Hamiltonian vector field, then
    \begin{equation}
	   \mathcal{L}_{X_\cH} \Omega = - (n+1) \reeb(\cH) \Omega, 
    \end{equation}
\end{theorem}

\begin{remark}
    From equations~\eqref{eqn:defHVF} it is possible to show also that for all contactomorphisms are of Hamiltonian nature~\cite{Bravetti2017, Geiges_2008}. Furthermore, the strict contactomorphism related to the Hamiltonian functions satisfying $\reeb(\cH) = 0$.
\end{remark}

Moreover, from equations~\eqref{eqn:defHVF} it is possible to recover one of the main differences with respect to the symplectic setting, the fact that the Hamiltonian function is not necessarily a constant of motion of its dynamics.
\begin{Prop}[\cite{Bravetti2017}]
    Let $(M,\eta,\cH)$ be a contact Hamiltonian system, with Hamiltonian possibly dependent on the time, then
    \begin{equation}
        \dot{\cH} = - \cH \reeb(\cH) + \partial_t \cH.
    \end{equation}
\end{Prop}
The autonomus Hamiltonian functions are conserved only in special cases: if $\reeb(\cH) = 0$ everywhere, the dynamics reduces locally to the symplectic case, where the Hamiltonian is conserved as usual~\cite{Bravetti2017}, and in the points where  $\cH = 0$, the level set is an invariant set. More generally, for time-dependent Hamiltonians, conservation holds precisely when
\begin{equation}
    \partial_t \cH = \cH\,\reeb(\cH).
\end{equation}

Theorem~\ref{thm:scalingvolume} and Proposition~\ref{prop:volumetransf} can be used to recover the contact Liouville equation~\cite{Bravetti_2015}, if a time-dependent probability density $\rho \in L^1_{\eta}(M)$ is transported by the flow of $X_H$, its density relative to the contact volume satisfies
\begin{equation}\label{eq:cLiouville}
\partial_t\rho=-X_\cH\rho+(n+1)\reeb(\cH)\rho,
\end{equation}
then the contact Liouville operator is given by $\hat{L}_\cH = - \imath X_\cH + \imath (n+1) \reeb(\cH)$.
\subsubsection{Jacobi Brackets}
The contact structure induces naturally a bracket structure, called Jacobi structure. The Jacobi structure provides an isomorphism of Lie algebras between the Lie bracket algebra of Hamiltonian vector fields and the Jacobi brackets of functions defined on the manifold~\cite{ZadraBravettiSeri2023FlowMethod}. This is a big difference with respect to the symplectic situation where Hamiltonian functions that differ form a constant gives rise to the same Hamiltonian vector field.

\begin{Def}~\cite{deLeonLainz2019,ZadraBravettiSeri2023FlowMethod}
    \label{def:Jacobi} Let $(M,\eta)$ a contact manifold, the
    induced Jacobi bracket is defined by:
    \begin{equation}
        \{ f, g\}_\eta := -\eta([X_f,X_g]).
    \end{equation}
\end{Def}
It is possible to define the Jacobi brakets as an independent setting of a contact manifold~\cite{Kirillov_1976, Lichnerowicz1978}, it relies on two objects: a skew-symmetric bi-vector field $\Lambda \in \mathfrak{X}(M) \wedge \mathfrak{X}(M)$, together with a vector field $E \in \mathfrak{X}(M)$ which satisfy:
    \begin{equation}
    \begin{dcases}
        [\Lambda,\Lambda]_{SN} = - 2 \Lambda \wedge E\\
        [\Lambda, E] = 0
    \end{dcases}
    \end{equation}
    The Jacobi brackets satisfy the Jacobi identity but they are not a derivation, i.e. they do not satisfies the Leibniz rule. On the contact manifolds this definition is realized by considering~\cite{deLeonLainz2020} $E = - \reeb$ and
    \begin{equation}
        \Lambda(d\eta) = 1, \qquad \Lambda(\eta, \cdot) = 0.
    \end{equation}
    
In the following proposition we collect some of the properties of the Jacobi brackets on a contact manifolds

\begin{Prop}~\cite{deLeonLainz2019,deLeonLainz2020} %
    \label{prop:propositionlemma}
    Let $(M,\eta)$ is a contact manifold then the induced Jacobi brackets defined in definition~\ref{def:Jacobi} satisfies the following relations:
    \begin{enumerate}
        \item $\{f,g\}_\eta = X_f g + g \reeb(f)$
        \item $\{f,1\}_\eta = \reeb(f)$
        \item $\{ f g, h\}_\eta = f \{g,h\}_\eta + g \{f,h\}_\eta + g f \reeb(h)$
    \end{enumerate}
\end{Prop}

\subsubsection{Contact Hamiltonian Integrability}
With the bracket structure it is then possible to define Hamiltonian integrability, and a complete intergable contact Hamiltonian system. In literature the definitions are different~\cite[Appendix B]{BanyagaHouenou2016}~\cite{Lerman_2001,Boyer_2011,ColomboDeLeonLainzLopezGordon2023} and it is still an active area of research~\cite{ColomboDeLeonLainzLopezGordon2023, ZadraSeri2026SymmetriesJMP}. In this work, we will refer to the following definition from~\cite{ZadraSeri2026SymmetriesJMP}.
\begin{Def}
    \label{def:CCHS}
    A complete contact integrable system $(M,\eta,\{f_0,\ldots, f_n\})$ is a triple formed by, a contact manifold $(M,\eta)$ and $n+1$ functions which satisfy:
    \begin{itemize}
        \item $\{f_j, f_k\}_\eta = 0$ for all $j,k \in \{0,\ldots,n\}$,
        \item $\beta = \sum_{j=0}^{n} \left(-1 \right)^{j n} f_j \bigwedge_{r=1}^{n} df_{j+r \pmod{n+1}} \neq 0$ almost everywhere.
    \end{itemize}
\end{Def}
The first condition is analogous to Poisson commutativity, while the second one represents linear independence almost everywhere on $(M,\eta)$ of the Hamiltonian vector fields generates by the Hamiltonian functions $(f_0, \ldots, f_n)$.
Moreover, the bracket structure characterizes also the conserved quantities~\cite{deLeonLainz2019, deLeonLainz2020,thesisZadra}, e.g. if on $(M,\eta)$ three functions $f,g,h$ are in involution then their ratio is a conserved quantity:
\begin{equation}
    X_f \left(\frac{f}{g}\right) = X_g \left(\frac{f}{g}\right) =X_h \left(\frac{f}{g}\right) =0.
\end{equation}

\subsubsection{Darboux theorem and local coordinates} 
In strict analogy to symplectic geometry, there exists a contact Darboux theorem~\cite{Geiges_2008,Bravetti2017}, which identify globally the canonical coordinates $(q^\mu, p_\mu, s)$, which permits to express in coordinates all the previous introduced objects:
\begin{align}
	\eta =& dS - p_i d q^i, \\
	\reeb =& \frac{\partial}{\partial s}, \\
	\Omega =& dS \wedge dq^1 \wedge dp_1 \wedge \dots \wedge dq^n \wedge dp_n.
\end{align}
The contact dynamics in canonical coordinates look like:
\begin{equation}
\begin{dcases}
	\dot{q}^i = \frac{\partial \cH}{\partial p_i}, \\
	\dot{p}_i = - \frac{\partial \cH}{\partial q^i} - p_i\frac{\partial \cH}{\partial s}, \\
	\dot{s} \,= p_i \frac{\partial \cH}{\partial p_i} - \cH.
\end{dcases}
\end{equation}
this last equation shows that the canonical coordinate $s$ can be interpreted as the ``action'' of the system~\cite{Bravetti2017}.

%%%%%%%%%%%%%%%%%%%%%%%%%%%%%%%%%%%%%%%%%%%%%%%%%%%%%%%%%%%%%%%%%%%%%%%%%%%%%%%%%%%%%%%%%%%%%%%%%%%%%%%%%%
\section{Contact Koopman von Neumann formulation}
\label{sec:contacKoopman}
%%%%%%%%%%%%%%%%%%%%%%%%%%%%%%%%%%%%%%%%%%%%%%%%%%%%%%%%%%%%%%%%%%%%%%%%%%%%%%%%%%%%%%%%%%%%%%%%%%%%%%%%%%
In this section, we introduce and characterize the Koopman von Neumann formulation in the contact setting.
The contact postulates are the really same of the symplectic setting, with just a couple of difference: the Hilbert space now is over a contact manifold, the operators that commutes are $\hat{Q}$, $\hat{P}$ and $\hat{S}$ which act as multiplicative operators, but the most crucial difference is on the operator generating the dynamics.

\begin{Def}
    The contact Koopman-von Neumann evolution operator associated to the contact Hamiltonian system $(M,\eta, \cH)$, is
    \begin{equation}
        K_{\cH} := - \imath X_\cH + \imath \frac{n+1}{2}\reeb(\cH) \mathbb{I},
    \end{equation}
    and the evolution equation takes the form:
    \begin{equation}
        \imath \partial_t \psi(t,x) = K_\cH \psi(t,x).\label{eqn:cKvNpde}
    \end{equation}
    The couple $(L^2(M,\eta), \hat{K}_\cH)$ is a contact Koopman von Neumann (cKvN) system.
\end{Def}

It is immediate to see how the cKvN operator is different from the contact Liouville operator, and their difference relies on the presence of the divergence term, compatible with the actions on half densities. 

\begin{theorem}
    \label{thm:KvNchar}
    Let $(M,\eta,\cH)$ be a contact Hamiltonian system, where $(M,\eta)$ is a contact manifold without boundary, and $(L^2(M,\eta), \hat{K}_\cH)$ the correspondent contact KvN system, then: 
    \begin{enumerate}[a)]
        \item The $\rho(t,x) = \lvert \psi(t,x) \rvert^2$ and
        \begin{equation}
            \imath \partial_t \psi(t,x) = \hat{K}_\cH \psi(t,x),
        \end{equation}
        with $\hat{K}_\cH$ self-adjoint, then $\rho(t,x)$ satisfies the contact Liouville equation. 
        %%%%%%%%%%%%%%%%%%%%%%%%%%%%%%%%%%%%%%%%%%%%%%%%%%%%%%%%%%
        \item The operator $\hat{K}_\cH$ is symmetric on $C_c^\infty(M)\subset L^2(M,\eta)$, with respect to the standard inner product
        \begin{equation}
            \langle \phi,\psi\rangle= \int_M \overline{\phi}\, \psi\,\Omega_M.
        \end{equation}
        %%%%%%%%%%%%%%%%%%%%%%%%%%%%%%%%%%%%%%%%%%%%%%%%%%%%%%%%%%
        \item If the contact Hamiltonian vector field $X_\cH$ is complete, then the operator $\hat{K}_\cH$ defined on $C_c^\infty(M)\subset L^2(M,\eta)$ is essentially self-adjoint. Its closure generates a strongly continuous one-parameter unitary group.
        %%%%%%%%%%%%%%%%%%%%%%%%%%%%%%%%%%%%%%%%%%%%%%%%%%%%%%%%%%
        \item Given any two Hamiltonian functions $f,g$ then:
        \begin{equation}
            [\hat{K}_f,\hat{K}_g] = -\imath\hat{K}_{\{f,g\}_\eta},
        \end{equation}
        so it is an isomorphism of Lie algebra. Moreover, if $f,g$ are two Jacobi commuting functions than also $\hat{K}_f$ and $\hat{K}_g$ commute.
        %%%%%%%%%%%%%%%%%%%%%%%%%%%%%%%%%%%%%%%%%%%%%%%%%%%%%%%%%
        \item Let $\phi:(M,\eta)\to(M,\eta')$ be a contactomorphism such that $\phi^*\eta'=f_\phi\eta$, with $f_\phi>0$, and let $h=(f_\phi\cH)\circ\phi^{-1}$ be the transformed Hamiltonian. Then the unitary transformation $U_\phi:L^2(M,\eta)\to L^2(M,\eta')$, defined by
        \begin{equation}
        (U_\phi\psi)(y)=f_\phi(\phi^{-1}(y))^{-\frac{n+1}{2}}\psi(\phi^{-1}(y)),
        \end{equation}
        satisfies
        \begin{equation}
            U_\phi\hat{K}_\cH U_\phi^{-1}=\hat{K}'_h
        \end{equation}
        on $C_c^\infty(M)$. If $X_\cH$ is complete, the identity also holds for the corresponding self-adjoint closures, and the diagram in Figure~\ref{fig:diagramL2} is commutative.
    \end{enumerate}
\end{theorem}
\begin{figure}[t]
    \centering
    \begin{tikzcd}
	{(M,\eta,\mathcal{H})} && {(M,\eta',\mathcal{h})} \\
	\\
	{\left(L^2(M,\eta),\hat{K}_\mathcal{H}\right)} & {} & {\left(L^2(M,\eta'),\hat{K}_h\right)}
	\arrow["\phi", from=1-1, to=1-3]
	\arrow[from=1-1, to=3-1]
	\arrow[from=1-3, to=3-3]
	\arrow["{U_\phi}", from=3-1, to=3-3]
\end{tikzcd}\
        \caption{Commutative diagram}
        \label{fig:diagramL2}
\end{figure}

Before the prove of the theorem three computational Lemmas are necessary. The first one relates the relation between the Jacobi brackets and the Reeb vector field.
\begin{Lemma}
\label{prop:reebcomm}
    \begin{equation}
        \reeb(\{F,G\}_\eta) = X_F(\reeb(G)) - X_G(\reeb(F)) 
    \end{equation}
\end{Lemma}
\begin{proof}
    By using Proposition~\ref{prop:propositionlemma} it follows that
    \begin{align}
        \reeb(\{F,G\}_\eta) &= \{-1, \{F,G\}_\eta\}_\eta\\
        &= - \{ G,\{-1,F\}\} - \{F,\{G,-1\}\} \\
        &= - \{ G, \reeb(F) \} - \{F,- \reeb(G)\}\\
        &= - X_G(\reeb(F)) - \reeb(F) \reeb(G) + X_F(\reeb(G)) -\reeb(G) \reeb(F) \\
        &= X_F(\reeb(G)) - X_G(\reeb(F)).
    \end{align}
    Consequently if two Hamiltonian functions are in involution, then $X_F(\reeb(G)) = X_G(\reeb(F))$.
\end{proof}
The second Lemma presents the transformation law under contactomorphism of the contraction between the differential of the contact Hamiltonian function and the Reeb vector field.

\begin{Lemma}
    \label{lemma:proptransf}
    Let $\phi$ be a contactomorphism of $(M,\eta)$, i.e. $
        \phi^*\eta = f_\phi\,\eta$, for some nowhere-vanishing function $f_\phi$, and let $\reeb$ and $\reeb'$ denote the Reeb vector fields of $\eta$ and $\phi^*\eta$ respectively, so that \begin{equation}
        \phi^{-1}_*\reeb' = X_{-\frac{1}{f_\phi}}.
    \end{equation}
    Given a Hamiltonian function $H$ on $M$, define the transformed Hamiltonian
    \begin{equation}
        h := (f_\phi H)\circ \phi^{-1}.
    \end{equation}
    Then the contraction of $\reeb'$ with $h$ transforms as
    \begin{equation}
        \reeb'(h) = \left(\reeb(H) - \frac{X_H(f_\phi)}{f_\phi}\right)\circ \phi^{-1}.
    \end{equation}
    In particular, if $\phi$ is a strict contactomorphism ($f_\phi \equiv 1$), then $\reeb' = \phi_*\reeb$ and $\reeb'(h) = \reeb(H)\circ\phi^{-1}$, i.e. the Hamiltonian's evolution along the Reeb direction is preserved.
\end{Lemma}
\begin{proof}
    \begin{align}\reeb' (h) &= \reeb' (f_\phi \, H \circ \phi^{-1})\\
    &= d (f_\phi H \circ \phi^{-1})(\reeb') \\
& = ({\phi^{-1}}^* d (f_\phi H) ) (\reeb') \\
& = d (f_\phi H) (\phi^{-1}_* \reeb') \circ \phi^{-1}\\
& = d (f_\phi H) (X_{-\frac{1}{f_\phi}}) \circ \phi^{-1}\\
& = d (f_\phi H) \left(\frac{1}{f_\phi^2} X_{f_\phi} + \frac{2}{f_\phi} \reeb \right) \circ \phi^{-1}\\
& = \frac{1}{f_\phi^2}\left( H X_{f_\phi}(f_\phi) + f_\phi X_{f_\phi}(H) \right) + \frac{2}{f_\phi}\left( H \reeb(f_\phi) + f_\phi \reeb(H) \right) \circ \phi^{-1}\\
& = \left(- \frac{H \reeb(f_\phi)}{f_\phi} + \frac{1}{f_\phi} X_{f_\phi}(H) + \frac{2 H \reeb(f_\phi)}{f_\phi} + 2 \reeb(H) \right) \circ \phi^{-1}\\
& = \left( \frac{1}{f_\phi} X_{f_\phi}(H) + \frac{H \reeb(f_\phi)}{f_\phi} + 2 \reeb(H) \right) \circ \phi^{-1} \\
& = \left( - \frac{1}{f_\phi} \left(- X_{f_\phi}(H) - H \reeb(f_\phi) \right) + 2 \reeb(H) \right) \circ \phi^{-1} \\
& = \left( 2 \reeb(H) - \frac{1}{f_\phi}\left\{H ,f_\phi \right\} \right) \circ \phi^{-1} \\
& = \left(\reeb(H) - \frac{X_H(f_\phi)}{f_\phi}\right) \circ \phi^{-1}
\end{align}
\end{proof}

The last lemma provides the general solution of the cKvN equation and the contact Liouville equation, which gives the unitary evolution operator.

\begin{Lemma}\label{prop:generalsolution}
Let $\phi^t$ denote the flow generated by the vector field $X_{\cH}$. Consider the Cauchy problem
\begin{equation}
\begin{dcases}
\partial_t f+X_{\cH}(f)=k\,\reeb(\cH)f,\\
f(t_0,x)=f_0(x),& x\in M,
\end{dcases}
\end{equation}
with $f:I\times M \to \mathbb{R}$ and $k \in \mathbb{R}$. Then, if the flow $\phi^t$ is defined on $I$, the Cauchy problem admits a unique solution, given by
\begin{equation}
\label{eq:generalsolution}
f(t,x)= f_0\!\left(\phi^{t_0-t}(x)\right) \exp\!\left( k\int_{t_0}^{t} \reeb(\cH)\!\left(\phi^{\xi-t}(x)\right) \,d\xi \right).
\end{equation}
\end{Lemma}

\begin{proof}
Let $\gamma(\xi) = \phi^{\xi-t}(x)$ be the characteristic curve passing through $x$ at time $\xi = t$. Note that $\gamma(t) = x$ and $\gamma(t_0) = \phi^{-t+t_0}(x)$. Define the pulled-back function along this characteristic as $g(\xi) = f(\xi, \gamma(\xi))$. Computing the total derivative with respect to the parameter $\xi$ yields:
\begin{equation}
    \frac{d}{d\xi} g(\xi) = \partial_\xi f(\xi, \gamma(\xi)) + \nabla f(\xi, \gamma(\xi)) \cdot \frac{d\gamma(\xi)}{d\xi}
\end{equation}
Since $\frac{d\gamma(\xi)}{d\xi} = X_\cH(\xi, \gamma(\xi))$, we exploit the governing PDE to substitute the terms:
\begin{equation}
    \frac{d}{d\xi} g(\xi) = \partial_\xi f(\xi, \gamma(\xi)) + X_\cH f(\xi, \gamma(\xi)) = k \, \reeb(\cH)(\xi, \gamma(\xi)) g(\xi)
\end{equation}
This establishes a first-order linear ordinary differential equation. Integrating from $\xi = t_0$ to $\xi = t$:
\begin{equation}
    g(t) = g(t_0) \exp \left( k \int_{t_0}^t \reeb(\cH)(\xi, \gamma(\xi)) d\xi \right)
\end{equation}
Reverting to the definition of $g(\xi)$ and enforcing the boundary initial values yields the claim:
\begin{equation}
    f(t,x) = f_0\left(\phi^{-t+t_0}(x)\right) \exp \left( k \int_{t_0}^t \reeb(\cH)\left(\xi, \phi^{\xi-t}(x)\right) d\xi \right)
\end{equation}
\end{proof}

\begin{proof}~[Theorem~\ref{thm:KvNchar}]
\leavevmode\par
    \begin{enumerate}[a)] 
        \item By directly computing and using the self-adjointness of $K_\cH$:
        \begin{align}
            \partial_t|\psi|^2
            &= -\imath\,\overline{\psi}\,\hat K_\cH\psi
            +\imath\,\psi\,\overline{\hat K_\cH\psi}\\
            &= -X_\cH(|\psi|^2)
            +(n+1)\reeb(\cH)|\psi|^2.
        \end{align}
        which coincide with equation~\eqref{eq:cLiouville}.
        %%%%%%%%%%%%%%%%%%%%%%%%%%%%%%%%%%%%%%%%%%%%%%%%%
        %%%%%%%%%%%%%%%%%%%%%%%%%%%%%%%%%%%%%%%%%%%%%%%%%
        \item Let $\psi,\chi\in C_c^\infty(M)$. We want to prove that
        \begin{equation}
            \langle \hat{K}_\cH\psi,\chi\rangle = \langle \psi,\hat{K}_\cH\chi\rangle.
        \end{equation}
        Since $\psi$ and $\chi$ are compactly supported, the differential form $\overline{\psi}\chi\,\Omega_M$ is also compactly supported. Therefore, since $M$ has no boundary, Stokes' theorem gives
        \begin{align}
            0 &= \int_M \Lie_{X_\cH}\left(\overline{\psi}\chi\,\Omega_M\right) \\
            &= \int_M X_\cH(\overline{\psi}\chi)\,\Omega_M + \int_M \overline{\psi}\chi\,\Lie_{X_\cH}\Omega_M \\
            &= \int_M \left(\overline{X_\cH\psi}\chi + \overline{\psi}X_\cH\chi -(n+1)\reeb(\cH)\overline{\psi}\chi\right)\Omega_M,
        \end{align}
        where we used Theorem~\ref{thm:scalingvolume}. Consequently,
        \begin{equation}
            \int_M \overline{X_\cH\psi}\chi\,\Omega_M = -\int_M \overline{\psi}X_\cH\chi\,\Omega_M +(n+1)\int_M \reeb(\cH)\overline{\psi}\chi\,\Omega_M.
        \end{equation}
        We can now compute
        \begin{align}
            \langle \hat{K}_\cH\psi,\chi\rangle &= \left\langle \left(-\imath X_\cH +\imath\frac{n+1}{2}\reeb(\cH)\mathbb{I}\right)\psi,\chi\right\rangle \\
            &= \imath\int_M \overline{X_\cH\psi}\chi\,\Omega_M -\imath\frac{n+1}{2}\int_M \reeb(\cH)\overline{\psi}\chi\,\Omega_M \\
            &= -\imath\int_M \overline{\psi}X_\cH\chi\,\Omega_M +\imath(n+1)\int_M \reeb(\cH)\overline{\psi}\chi\,\Omega_M -\imath\frac{n+1}{2}\int_M \reeb(\cH)\overline{\psi}\chi\,\Omega_M \\
            &= -\imath\int_M \overline{\psi}X_\cH\chi\,\Omega_M +\imath\frac{n+1}{2}\int_M \reeb(\cH)\overline{\psi}\chi\,\Omega_M \\
            &= \int_M \overline{\psi}\left(-\imath X_\cH\chi +\imath\frac{n+1}{2}\reeb(\cH)\chi\right)\Omega_M \\
            &= \langle \psi,\hat{K}_\cH\chi\rangle.
        \end{align}
        Therefore, $\hat{K}_\cH$ is symmetric on $C_c^\infty(M)$.
        %%%%%%%%%%%%%%%%%%%%%%%%%%%%%%%%%%%%%%%%%%%%%%%%%%%
        %%%%%%%%%%%%%%%%%%%%%%%%%%%%%%%%%%%%%%%%%%%%%%%%%%%
        \item We prove the essential self-adjointness of $\hat{K}_\cH$ using Theorem VIII.10 of~\cite{ReedSimonI}. This theorem states that, given a strongly continuous one-parameter unitary group $U(t)$ and a dense subspace $D$ invariant under $U(t)$, if $U(t)$ is strongly differentiable on $D$, then the operator
        \begin{equation}
            B\psi=\imath\left.\frac{d}{dt}U(t)\psi\right|_{t=0},\qquad \psi\in D,
        \end{equation}
        is essentially self-adjoint, and its closure generates $U(t)=e^{-\imath t\overline{B}}$.
        
        Let $\phi^t$ denote the flow generated by $X_\cH$. Since $X_\cH$ is complete, $\phi^t$ is defined for every $t\in\mathbb{R}$ and is a diffeomorphism of $M$. Theorem~\ref{thm:scalingvolume} implies
        \begin{align}
            \frac{d}{dt}(\phi^t)^*\Omega_M &= (\phi^t)^*(\Lie_{X_\cH}\Omega_M) \\
            &= -(n+1)(\reeb(\cH)\circ\phi^t)(\phi^t)^*\Omega_M,
        \end{align}
        and therefore
        \begin{equation}
            \label{eqn:cflowvolume}
            (\phi^t)^*\Omega_M = \exp\left(-(n+1)\int_0^t \reeb(\cH)(\phi^\xi(x))\,d\xi\right)\Omega_M.
        \end{equation}
        
        Using Lemma~\ref{prop:generalsolution}, we define the evolution operator $U(t)$ by
        \begin{equation}
            \label{eqn:cunitaryev}
            (U(t)\psi)(x) = \psi(\phi^{-t}(x))\exp\left(\frac{n+1}{2}\int_0^t \reeb(\cH)(\phi^{\xi-t}(x))\,d\xi\right).
        \end{equation}
        
        To prove the unitarity of $U(t)$, we consider the change of variables $x=\phi^t(y)$. Using equations~\eqref{eqn:cflowvolume} and~\eqref{eqn:cunitaryev}, we obtain
        \begin{align}
            \|U(t)\psi\|^2 &= \int_M |U(t)\psi(x)|^2\,\Omega_M \\
            &= \int_M |\psi(\phi^{-t}(x))|^2 \exp\left((n+1)\int_0^t \reeb(\cH)(\phi^{\xi-t}(x))\,d\xi\right)\Omega_M \\
            &= \int_M |\psi(y)|^2 \exp\left((n+1)\int_0^t \reeb(\cH)(\phi^\xi(y))\,d\xi\right)(\phi^t)^*\Omega_M \\
            &= \int_M |\psi(y)|^2\,\Omega_M = \|\psi\|^2.
        \end{align}
        Moreover, the group property of the flow and the additivity of the integral imply that $U(t+s)=U(t)U(s)$ and $U(0)=\mathbb{I}$. Since $U(-t)=U(t)^{-1}$, each $U(t)$ is unitary.

        The next step is strong continuity: let $\psi\in C_c^\infty(M)$ and be $K$ its support, for any $\delta>0$, the set $K_\delta=\bigcup_{|t|\leq\delta}\phi^t(K)$ is also compact and $\operatorname{supp}(U(t)\psi)\subseteq K_\delta$ for every $|t|\leq\delta$.
        Since $\reeb(\cH)$ is smooth, there exists $C_\delta>0$ such that $|\reeb(\cH)|\leq C_\delta$ on $K_\delta$. Consequently,
        \begin{equation}
            |U(t)\psi(x)-\psi(x)|^2
            \leq 4\left(\sup_{y\in K_\delta}|\psi(y)|\right)^2
            e^{(n+1)C_\delta\delta}\mathbf{1}_{K_\delta}(x),
            \qquad |t|\leq\delta.
        \end{equation}
        The right-hand side is independent of $t$ and integrable, since $K_\delta$ has finite measure with respect to $\Omega_M$. Furthermore, the smoothness of the flow implies that $U(t)\psi(x)\to\psi(x)$ pointwise as $t\to0$. Therefore, the dominated convergence theorem~\cite[Theorem 2.24]{Folland1999} gives
        \begin{equation}
            \lim_{t\to0}\|U(t)\psi-\psi\|=0.
        \end{equation}
        Since $C_c^\infty(M)$ is dense in $L^2(M,\eta)$ and $U(t)$ is unitary, strong continuity extends to every $\psi\in L^2(M,\eta)$.
        
        It remains to verify the invariance and strong differentiability required by Theorem VIII.10. Since $\phi^t$ is a diffeomorphism and the exponential factor in~\eqref{eqn:cunitaryev} is smooth, we have
        \begin{equation}
            U(t)C_c^\infty(M)=C_c^\infty(M),\qquad t\in\mathbb{R}.
        \end{equation}
        Furthermore, for every $\psi\in C_c^\infty(M)$, differentiation of~\eqref{eqn:cunitaryev} at $t=0$ gives
        \begin{align}
            \left.\frac{d}{dt}U(t)\psi\right|_{t=0} &= -X_\cH\psi+\frac{n+1}{2}\reeb(\cH)\psi = -\imath\hat{K}_\cH\psi.
        \end{align}
        The derivative exists in $L^2$ by smoothness, compact support and dominated convergence. Strong differentiability for all $t\in\mathbb{R}$ then follows from the group property.
        
        Therefore, $C_c^\infty(M)$ is a dense invariant subspace on which $U(t)$ is strongly differentiable, and
        \begin{equation}
            \hat{K}_\cH\psi=\imath\left.\frac{d}{dt}U(t)\psi\right|_{t=0}.
        \end{equation}
        All the hypotheses of Theorem VIII.10 are satisfied. Consequently, $\hat{K}_\cH$ is essentially self-adjoint on $C_c^\infty(M)$, and its self-adjoint closure $\overline{\hat{K}_\cH}$ generates the unitary group
        \begin{equation}
            U(t)=e^{-\imath t\overline{\hat{K}_\cH}}.
        \end{equation}
    %%%%%%%%%%%%%%%%%%%%%%%%%%%%%%%%%%%%%%%%%%%%%%%%%%%
    %%%%%%%%%%%%%%%%%%%%%%%%%%%%%%%%%%%%%%%%%%%%%%%%%%%
    \item \label{thm:KvncharLie}The isomomorphism of Lie algebras can be proved by the explicit computation:
        \begin{align}
        \left[\hat{K}_\cH, \hat{K}_F\right] \phi &= \left[- \imath X_\cH +\imath \frac{n+1}{2} \reeb(\cH) , -\imath X_F + \imath \frac{n+1}{2} \reeb(F) \right] \\
        &= \left[-\imath X_\cH, -\imath X_F \right] + \frac{n+1}{2}\left( \left[i \reeb(\cH), - \imath X_F \right] +\left[-\imath X_\cH, \imath \reeb(F)  \right] \right) \\
        &= - X_{\{H,F\}} +  \frac{n+1}{2}\bigg( \reeb(\cH) X_F(\phi)) -X_F (\reeb(\cH)\phi) \\ & \qquad \qquad + X_\cH(\reeb(F)\phi) - \reeb(F) X_\cH(\phi) \bigg) \\
        &= - X_{\{H,F\}} +  \frac{n+1}{2}\bigg(\reeb(\cH) X_F(\phi) - X_F(\reeb(\cH)) \phi - \reeb(\cH) X_F(\phi) + \\
        & \qquad \qquad + X_\cH(\reeb(F)) \phi + \reeb(F) X_\cH(\phi) - \reeb(F) X_\cH(\phi) \bigg) \\
        & = - X_{\{H,F\}} +  \frac{n+1}{2} \left(X_\cH(\reeb(F)) - X_F(\reeb(\cH))  \right) \label{line:1}\\
        & = - X_{\{H,F\}} +  \frac{n+1}{2} \reeb(\{H,F\}) \label{line:2}\\
        & = - \imath \left( - \imath X_{\{H,F\}_\eta} + \imath  \frac{n+1}{2} \reeb(\{H,F\}) \right) \\
        & = - \imath \hat{L}_{\{\cH,F\}_\eta}.
        \end{align}
        Where, between equation~\eqref{line:1} and equation~\eqref{line:2}, we exploited Lemma~\ref{prop:reebcomm}. 
    %%%%%%%%%%%%%%%%%%%%%%%%%%%%%%%%%%%%%%%%%%%%%%%%%%%%%%
    %%%%%%%%%%%%%%%%%%%%%%%%%%%%%%%%%%%%%%%%%%%%%%%%%%%%%%
    \item Given a contactomorphism $\phi$, the unitary transformation of square-integrable functions follows from the transformation law of the contact volume form established in Proposition~\ref{prop:volumetransf}. In particular, for $\psi\in L^2(M,\eta')$, the inverse transformation is given by
    \begin{equation}
        \tilde{\psi}(x)=(U_\phi^{-1}\psi)(x)=\psi(\phi(x))f_\phi(x)^{\frac{n+1}{2}}.
    \end{equation}
    Indeed, using $\phi^*\Omega'_M=f_\phi^{n+1}\Omega_M$, we obtain
    \begin{align}
        \|U_\phi\psi\|^2_{L^2(M,\eta')} &= \int_M |U_\phi\psi(y)|^2\,\Omega'_M \\
        &= \int_M |\psi(x)|^2 f_\phi(x)^{-(n+1)}\phi^*\Omega'_M \\
        &= \int_M |\psi(x)|^2\,\Omega_M = \|\psi\|^2_{L^2(M,\eta)}.
    \end{align}
    The transformation is therefore unitary.

    We now apply $\hat{K}_\cH$ to $\tilde{\psi}$. Using the definition of the contact KvN operator, we obtain
    \begin{align}
        \hat{K}_\cH\tilde{\psi}(x)
        &= -\imath X_\cH\left(\psi(\phi(x))f_\phi(x)^{\frac{n+1}{2}}\right)+\imath\frac{n+1}{2}\reeb(\cH)\psi(\phi(x))f_\phi(x)^{\frac{n+1}{2}} \\
        &= -\imath f_\phi(x)^{\frac{n+1}{2}}X_\cH(\psi\circ\phi)(x) -\imath\frac{n+1}{2}f_\phi(x)^{\frac{n-1}{2}}X_\cH(f_\phi)(x)\psi(\phi(x)) \\
        &\qquad +\imath\frac{n+1}{2}\reeb(\cH)f_\phi(x)^{\frac{n+1}{2}}\psi(\phi(x)).
    \end{align}
    Since $\phi$ maps the Hamiltonian vector field $X_\cH$ into $X'_h$, where $h=(f_\phi\cH)\circ\phi^{-1}$, we have
    \begin{equation}
    X_\cH(\psi\circ\phi)=(X'_h\psi)\circ\phi.
    \end{equation}
    Consequently,
    \begin{align}
        \hat{K}_\cH\tilde{\psi}(x)
        &= f_\phi(x)^{\frac{n+1}{2}}\bigg[-\imath(X'_h\psi)(\phi(x))+\imath\frac{n+1}{2}\left(\reeb(\cH)(x)-\frac{X_\cH(f_\phi)(x)}{f_\phi(x)}\right)\psi(\phi(x))\bigg].
    \end{align}
    Applying $U_\phi$ to both sides and using its definition, we obtain
    \begin{align}
        (U_\phi\hat{K}_\cH U_\phi^{-1}\psi)(y)
        &= -\imath X'_h\psi(y) +\imath\frac{n+1}{2}\left(\reeb(\cH)-\frac{X_\cH(f_\phi)}{f_\phi}\right)\circ\phi^{-1}(y)\,\psi(y).
    \end{align}
    Finally, by Lemma~\ref{lemma:proptransf}, the transformation law of the Reeb derivative is
    \begin{equation}
        \reeb'(h)=\left(\reeb(\cH)-\frac{X_\cH(f_\phi)}{f_\phi}\right)\circ\phi^{-1}.
    \end{equation}
    Substituting this relation into the previous expression yields
    \begin{align}
        (U_\phi\hat{K}_\cH U_\phi^{-1}\psi)(y) = -\imath X'_h\psi(y)+\imath\frac{n+1}{2}\reeb'(h)(y)\psi(y) = \hat{K}'_h\psi(y).
    \end{align}
    Therefore $U_\phi\hat{K}_\cH U_\phi^{-1}=\hat{K}'_h$. Since $U_\phi$ maps $C_c^\infty(M)$ onto itself, the identity holds on this domain. Moreover, when $X_\cH$ is complete, the operators are essentially self-adjoint by point (c), and the identity extends to their self-adjoint closures.
    \end{enumerate}
\end{proof}

\begin{remark}
    For a contact Hamiltonian system $(M,\eta, \cH)$ a sufficient and necessary condition for the completeness for contact Hamiltonian flow doesn't exist in literature. A sufficient condition~\cite[Proposition 4.6]{crasmareanu2011completeness} requires the existence of a proper function $g \in C^{\infty}(M)$ and positive real constants $a,b$, which satisfy
    \begin{equation}
        \lvert d\eta(X_f,X_g)(x) - f(x)\reeb(g(x)) \rvert \leq a \lvert g(x) \rvert + b,
    \end{equation}
    for each $x \in M$.
    A particular interesting example of non complete flow is given by the Hamiltonian system (see ~\cite[Example 2.1]{Liu_2022} and~\cite{Bravetti_2021}):
    \begin{equation}
        \left(\mathbb{R}^3, \, \eta = ds - p dq, \, \cH = \frac{p^2}{2} + \frac{q^2}{2} + \frac{s^2}{2} - \frac{1}{2} \right).
    \end{equation}
    The invariant surface corresponding to $\cH = 0$ is a sphere of unitary radius in the phase space. In this model there exist at least one non complete trajectory: given the initial condition $(q,p,s)=(0,0,s_0)$ with recover the equations of motion $(\dot{q} = 0, \dot{p} = 0, \dot{s} = -\frac{s^2}{2}+\frac12)$, which blow-up in finite time equal to $T = \log\left(\frac{-1 + s_0}{1+s_0} \right)$ with $s_0<-1$.
    Similar type of Hamiltonians are presents in applications in cosmology~\cite{Sloan2018}.
    Although the contact Hamiltonian flow is not complete, its only incomplete trajectories are the two orbits on the $s$-axis with $|s|>1$. Since these trajectories are a zero measure set, the associated cKvN evolution  define a strongly continuous one-parameter unitary group on $L^2(M,\eta)$.
\end{remark}

%%%%%%%%%%%%%%%%%%%%%%%%%%%%%%%%%%%%%%%%%%%%%%%%%%%%%%%%%%%%%%%%%%%%%%%%%%%%%%%%%%%%%%%%%%%%%%%%%%%%%%%%%%%%%%%%
\subsection{Differential Entropy}
A classical way to study the information carried by a wave function, or a probability distribution is the differential entropy~\cite{Jaynes1963} (sometimes called also Shannon entropy), which is the starting point of the entropic uncertainty relations~\cite{Biaynicki_Birula_1975}.

\begin{Def}[Entropy] Given a probability space $(\Omega, \mu)$, the entropy associated to a sufficietly regular probability distribution $\rho \in L^1(\Omega, \mu)$ is:
    \begin{equation}
    \Sigma[\rho] := \langle - \log(\rho)\rangle_{\eta} = -\int_\Omega \rho \log(\rho) d\mu.
\end{equation}
\end{Def}

\begin{theorem}
    \label{thm:entropycontact}
    Let $\rho \in L^1(M,\eta)$ be sufficiently regular probability distribution and $\cH$ a contact Hamiltonian that generates the contact Liouville equation:
    \begin{equation}
        \partial_t \rho = \hat{L}_\cH \rho,
    \end{equation}
    then
    \begin{equation}
        \frac{d}{dt} \Sigma(t) = -(n+1) \langle \reeb(\cH) \rangle_\eta.
    \end{equation}
    \end{theorem}
    \begin{proof}
    Since for Theorem~\ref{thm:KvNchar} the evolution of $\rho(t,x)$ is equivalent to the evolution of the square modulus of $\psi(t,x)$, we study the evolution of the first by the Liouville equation.
    \begin{align}
        \frac{d}{d t} \Sigma[\rho](t) &= -\int_M \left[\left(\partial_t \rho + X_\cH \rho \right)\left( 1 + \log \rho \right) - (n+1) \reeb(\cH) \rho \log \rho \right] \Omega_M \\
        & = -\int_M \left[ \left((n+1) \reeb(\cH) \rho \right) \left(1 + \log \rho \right) - (n+1) \reeb(\cH) \rho \log \rho \right] \Omega_M \\
        & = -(n+1) \int_M \rho\, \reeb(\cH) \Omega_M \\
        & = - (n+1) \langle \reeb(\cH) \rangle_\eta.
    \end{align}
\end{proof}

This is another difference in between contact and symplectic settings, in fact, over symplectic manifold $(S,\omega)$ the differential entropy is conserved as can be checked by:
\begin{equation}
\Sigma[\rho_S] = -\int_\Omega \rho_S \log(\rho_S) \, \Omega_S
\end{equation}
we compute the time derivative:
\begin{equation}
    \frac{d}{dt} \Sigma[\rho_S] = -\int (\partial_t \rho_S + \hat{L}\rho_S) (1+\log \rho) \,  \Omega_S = 0,
\end{equation}
since $(\partial_t \rho_S + \hat{L}\rho_S)$ vanishes from the standard Liouville equation.

%%%%%%%%%%%%%%%%%%%%%%%%%%%%%%%%%%%%%%%%%%%%%%%%%%%%%%%%%%%%%%%%%%%%%%%%%%%%%%%%%%%%%%%%%%%%%
\subsection{Spectral decomposition of cKvN systems}
In Lemma~\ref{prop:generalsolution} it has been shown that the method of characteristic is suitable for the solution of the Cauchy problem associated to a cKvN system. Nevertheless, the Hilbert space setting suggest to approach the same probelm by using the diagonalization of the operator $\hat{K}_\cH$. To achieve the diagonalization let us cosider a complete contact integrable system $(M,\eta,\{f_0,\ldots, f_n\})$ as in Definition~\ref{def:CCHS}, and define the set:
\begin{equation}
    M_{\text{reg}} = \left\{x \in M\, :\, \beta(x)\neq0  \right\}.
\end{equation}
The function
\begin{align}
    g = \sqrt{\sum_{j=0}^n f_j^2},
\end{align}
is (positive)-homogeneus of degree one of the $\{f_j\}_j$, than $\{g,f_j\}_\eta = 0$ for each $j$~\cite{thesisZadra}. Moreover, for each point in $x\in M_{\text{reg}}$ the function $g$ is strictly positive, and consequently the conserved quantities
\begin{equation}
    k_j = \frac{f_j}{g},
\end{equation}
are well defined and smooth in $M_{\text{reg}}$.
\begin{Prop}
    Let $(M,\eta,\{f_0,\ldots, f_n\})$ be a completely integrable contact Hamiltonian system, then \begin{equation}
        k=(k_0,\ldots,k_n):M_{\mathrm{reg}}\longrightarrow S^n
    \end{equation}
    is a smooth submersion. Moreover, $X_{f_i}(k_j)=0$ for every $i,j=0,\ldots,n$. Consequently, the level sets of $k$ are invariant under all the Hamiltonian flows.
\end{Prop}
\begin{proof}
    Since $g>0$ on $M_{\mathrm{reg}}$, the map $k=(k_0,\ldots,k_n)$ is smooth and satisfies $\sum_{j=0}^n k_j^2=1$. Moreover, denoting by $\mathrm{vol}_{S^n}$ the standard volume form on $S^n$, the independence condition of Definition~\ref{def:CCHS} gives
    \begin{equation}
        \beta=g^{n+1}k^*\mathrm{vol}_{S^n}.
    \end{equation}
    Since $\beta\neq0$ and $g>0$ on $M_{\mathrm{reg}}$, we have $\operatorname{rank}(dk)=n$, proving that $k$ is a smooth submersion onto its image.
    Since the functions $k_j$ are conserved, their differentials satisfy
    \begin{equation}
        dk_{\phi_i^t(x)}\circ d\phi_i^t|_x=dk_x,
    \end{equation}
    where $\phi_i^t$ is the flow of $X_{f_i}$. Since $g$ remains positive along the flow and $\phi_i^t$ is a local diffeomorphism, the rank of $dk$ is preserved. Consequently, $M_{\mathrm{reg}}$ and the level sets of $k$ are invariant under all the Hamiltonian flows, wherever defined.
\end{proof}
Using this proposition it is possible to prove the diagonalization statement.
\begin{theorem}\label{thm:diagonalization}
    Let $(M,\eta,\{f_0,\ldots,f_n\})$ be a completely integrable contact Hamiltonian system, and assume that all the Hamiltonian vector fields $X_{f_j}$ are complete. Let $\hat{k}_j$ denote multiplication by $k_j$ on $L^2(M,\eta)$. Then:
    \begin{enumerate}
        \item The operators $\hat{k}_j$ are bounded and self-adjoint, and strongly commute with the self-adjoint closures of $\hat{K}_{f_i}$ for every $i,j=0,\ldots,n$.
        \item There exists a direct-integral decomposition
        \begin{equation}\label{eq:KvN_direct_integral}
            L^2(M,\eta)\simeq\int_{k(M_{\mathrm{reg}})}^\oplus\mathcal{H}_\kappa\,d\nu(\kappa),
        \end{equation}
        where $\mathcal{H}_\kappa$ is the Hilbert space associated with the invariant level set $k^{-1}(\kappa)$. In this representation,
        \begin{equation}
            \overline{\hat{K}_{f_i}}\simeq\int_{k(M_{\mathrm{reg}})}^\oplus\hat{K}_{f_i,\kappa}\,d\nu(\kappa).
        \end{equation}
        \item The operators $\overline{\hat{K}_{f_0}},\ldots,\overline{\hat{K}_{f_n}}$ strongly commute and therefore admit a joint spectral decomposition.
    \end{enumerate}
\end{theorem}

\begin{proof}
    Since $|k_j|\leq1$, the multiplication operators $\hat{k}_j$ are bounded and self-adjoint. Let $U_i(t)$ denote the unitary group generated by $\overline{\hat{K}_{f_i}}$, whose existence follows from Theorem~\ref{thm:KvNchar}, point (c). Since the functions $k_j$ are conserved along the Hamiltonian flows, we have
    \begin{equation}
        U_i(t)\hat{k}_j=\hat{k}_jU_i(t),\qquad i,j=0,\ldots,n.
    \end{equation}
    Consequently, the spectral projections of the multiplication operators $\hat{k}_j$ commute with $U_i(t)$, establishing their strong commutativity with $\overline{\hat{K}_{f_i}}$.

    By the previous proposition, $k:M_{\mathrm{reg}}\to S^n$ is a smooth submersion. The coarea formula~\cite[Section 3.2]{Federer1969} allows us to decompose the contact volume form along its level sets, obtaining
    \begin{equation}
        L^2(M_{\mathrm{reg}},\eta)\simeq\int_{k(M_{\mathrm{reg}})}^\oplus L^2(k^{-1}(\kappa),d\mu_\kappa)\,d\nu(\kappa),
    \end{equation}
    where $\kappa\in k(M_{\mathrm{reg}})\subseteq S^n$ denotes a fixed value of the conserved quantities, $d\nu$ is a measure on $k(M_{\mathrm{reg}})$ and $d\mu_\kappa$ is the corresponding measure on the level set $k^{-1}(\kappa)$. Since the complement of $M_{\mathrm{reg}}$ has measure zero with respect to $\Omega_M$ by Definition~\ref{def:CCHS}, this decomposition extends to $L^2(M,\eta)$.

    Since the unitary groups $U_i(t)$ commute with every bounded function of $k$, they are decomposable with respect to this direct integral~\cite{Folland2016}. Consequently, their self-adjoint generators admit the decomposition~\cite{Folland1999}
    \begin{equation}
        \overline{\hat{K}_{f_i}}\simeq\int_{k(M_{\mathrm{reg}})}^\oplus\hat{K}_{f_i,\kappa}\,d\nu(\kappa),
    \end{equation}
    where the fiber operators $\hat{K}_{f_i,\kappa}$ are self-adjoint for almost every $\kappa$.

    Finally, by Theorem~\ref{thm:KvNchar}, point (d), Jacobi commutativity implies
    \begin{equation}
        [\hat{K}_{f_i},\hat{K}_{f_j}]=-\imath\hat{K}_{\{f_i,f_j\}_\eta}=0.
    \end{equation}
    Since the corresponding Hamiltonian vector fields are complete, their flows commute globally. Consequently, the associated unitary groups, defined by the weighted pullback along these flows, satisfy
    \begin{equation}
        U_i(t)U_j(s)=U_j(s)U_i(t),\qquad s,t\in\mathbb{R}.
    \end{equation}
    Hence, the self-adjoint operators $\overline{\hat{K}_{f_0}},\ldots,\overline{\hat{K}_{f_n}}$ strongly commute and admit a joint spectral decomposition.
\end{proof}

%%%%%%%%%%%%%%%%%%%%%%%%%%%%%%%%%%%%%%%%%%%%%%%%%%%%%%%%%%%%%%%%%%%%%%%%%%%%%%%%%%%%%%%%%%%%%
\subsection{cKvN formalism and symplectization}
Given a symplectic Hamiltonian system, as shown in Section~\ref{subsec:Liouville} and in~\cite{Koopman1931,v_Neumann_1932}, one can equivalently describe the dynamics using the standard Koopman-von Neumann (KvN) formulation. Moreover, the symplectization procedure~\cite{Arnold_1989} provides a way to associate a symplectic Hamiltonian system with a contact Hamiltonian system (see Appendix~\ref{app:symplectization} for more details). The aim of this section is to establish the relationship between the cKvN formulation and the symplectic KvN formulation through the symplectization procedure.

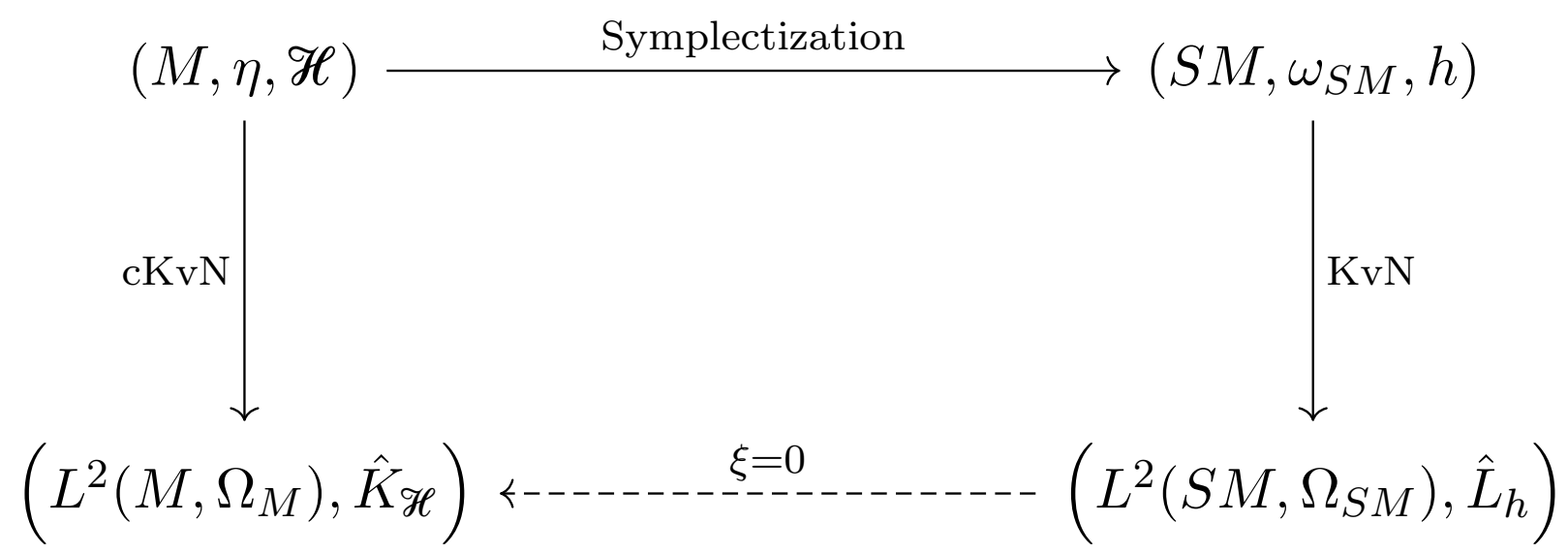
\begin{figure}[t]
    \centering
    \begin{tikzcd}[column sep=huge, row sep=large]
        {(M,\eta,\mathcal H)}
        &&
        {(SM,\omega_{SM},h)}
        \\
        \\
        {\left(L^2(M,\Omega_M),\hat{K}_{\mathcal H}\right)}
        &&
        {\left(L^2(SM,\Omega_{SM}),\hat{L}_h\right)}
        \arrow["{\mathrm{Symplectization}}", from=1-1, to=1-3]
        \arrow["{\mathrm{cKvN}}"', from=1-1, to=3-1]
        \arrow["{\mathrm{KvN}}", from=1-3, to=3-3]
        \arrow["{\xi=0}"', dashed, from=3-3, to=3-1]
    \end{tikzcd}
    \caption{Relation between the contact and symplectic Koopman--von Neumann formulations.}
    \label{fig:diagramL2symp}
\end{figure}

Given a contact Hamiltonian system $(M,\eta,\mathcal H)$, with dynamics generated by the contact Hamiltonian vector field $X_{\mathcal H}$, one can associate to it the symplectic Hamiltonian system
\begin{equation}
    \left(SM=\mathbb R\times M, \omega_{SM}=d(e^u\eta), h=e^u\mathcal H,\right)
\end{equation}
where $u$ is the coordinate on $\mathbb R$. The corresponding symplectic Hamiltonian vector field is given by
\begin{equation}
    Y_h = X_{\mathcal H} + \reeb(\mathcal H)\partial_u.
    \label{eq:symplectized_Hamiltonian_vector_field}
\end{equation}The symplectic Koopman-von Neumann formulation is provided as in the usual case illustrated in Section~\ref{sec:background}, that is a Hilbert space
\begin{equation}
    \left( L^2\left(M\times\mathbb R,\Omega_{SM}\right), \hat{L}_h\right),
\end{equation}
where $\Omega_{SM}$ is the standard symplectic volume~\eqref{eqn:sympvolume} and the Liouville operator
\begin{equation}
    \label{eq:Lh_original}
    \hat{L}_h=-\imath X_{\mathcal H}-\imath\reeb(\mathcal H)\partial_u,
\end{equation}
which gives the dynamics of the Liouville equation~\eqref{eqn:sympLiouville}. The relation between the two formulations is illustrated by the diagram in Figure~\ref{fig:diagramL2symp}. Starting from a contact Hamiltonian system, one can either construct its cKvN formulation directly or first apply symplectization and then the standard KvN formulation. The following proposition shows that these two procedures are related through a direct-integral decomposition of the symplectic KvN generator, in which the cKvN generator appears as the zero-frequency fiber. Thus, the diagram is understood at the level of operators, rather than as a unitary equivalence between the corresponding Hilbert spaces.

\begin{Prop}\label{prop:KvN_symplectization}
    Let $(M,\eta,\cH)$ be a contact Hamiltonian system whose Hamiltonian vector field $X_\cH$ is complete, and let $(SM,\omega_{SM},h)$ be its symplectization. Then there exists a unitary transformation
    \begin{equation}
         \mathfrak{U}:L^2(SM,\Omega_{SM})\longrightarrow\int_{\mathbb{R}}^\oplus L^2(M,\Omega_M)\,d\xi
    \end{equation}
    such that
    \begin{equation}
        \mathfrak{U}\overline{\hat L_h} \mathfrak{U}^{-1}=\int_{\mathbb{R}}^\oplus\overline{\hat L_{\cH,\xi}}\,d\xi,
        \label{eq:operator_direct_integral}
    \end{equation}
    where the operators
    \begin{equation}
        \hat L_{\cH,\xi}=\hat K_\cH+\xi \reeb(\cH),\qquad \xi\in\mathbb{R},
        \label{eq:fiber_operator}
    \end{equation}
    are essentially self-adjoint on $C_c^\infty(M)$, and $\reeb(\cH)$ acts multiplicatively. Moreover, $\hat L_{\cH,0}=\hat K_\cH$.
    \end{Prop}

\begin{proof}
    Since $SM = \mathbb{R} \times M$, there is an unitary transformation $O_u$~\cite{ReedSimonI}
    \begin{equation}
        \hat{O}_u:L^2(SM,\Omega_{SM})\longrightarrow L^2(\mathbb{R},du)\otimes L^2(M,\Omega_M),
    \end{equation}
    given by
    \begin{equation}
        (\hat{O}_u\Psi)(u,x)=e^{\frac{n+1}{2}u}\Psi(u,x).
        \label{eq:Ou}
    \end{equation}
    Indeed, using the expression of the symplectic volume form, we obtain
    \begin{equation}
        \|\hat O_u\Psi\|^2=\int_{\mathbb{R}\times M}e^{(n+1)u}|\Psi(u,x)|^2\,du\,\Omega_M=\|\Psi\|^2.
    \end{equation}
    Moreover, the transformation is invertible and then unitary. The action of $\hat{O}_u$ on the Liouville operator $\hat{L}_h$~\eqref{eq:Lh_original} is given by:
    \begin{align}
    \label{eq:Lh_transformed}
        \hat O_u\hat L_h\hat O_u^{-1}=-\imath X_\cH-\imath\reeb(\cH)\left(\partial_u-\frac{n+1}{2}\right)=\hat K_\cH-\imath\reeb(\cH)\partial_u.
    \end{align}
    Since the cKvN operator $\hat{K}_\cH$ does not depend on the variable $u$, it is possible to use the Fourier transform $\mathcal{F}_u$ as unitary transformation to obtain:
    \begin{equation}
        \mathcal{F}_u \left(\hat O_u\hat L_h\hat O_u^{-1}\right)\mathcal{F}_u^{-1}=\hat K_\cH+\xi M_{\reeb(\cH)},
    \end{equation}
    Since
    \begin{equation}
        L^2(\mathbb{R},d\xi)\otimes L^2(M,\Omega_M)\simeq\int_{\mathbb{R}}^\oplus L^2(M,\Omega_M)\,d\xi,
    \end{equation}
    we get the formal direct-integral decomposition of $\hat L_h$ into the operators $\hat L_{\cH,\xi}$.

    To establish the essential self-adjointness of the fiber operators, it can be use the same argument of Theorem~\ref{thm:KvNchar}, point (c). Let $U_0(t)$ denote the cKvN unitary group and define
    \begin{equation}
        U_\xi(t)=M_{e^{-\imath\xi I_t}}U_0(t),\qquad I_t(x)=\int_0^t\reeb(\cH)(\phi^{s-t}(x))\,ds.
    \end{equation}
    The identity $I_{t+s}(x)=I_t(x)+I_s(\phi^{-t}(x))$ implies that $U_\xi(t)$ is a strongly continuous one-parameter unitary group preserving $C_c^\infty(M)$, with infinitesimal generator $\hat L_{\cH,\xi}=\hat K_\cH+\xi M_{\reeb(\cH)}$. Therefore, the same argument as in Theorem~\ref{thm:KvNchar}, point (c), establishes essential self-adjointness on $C_c^\infty(M)$ for every $\xi\in\mathbb R$.

    Finally, completeness of $X_\cH$ implies completeness of the symplectized Hamiltonian vector field $Y_h$. Denoting by $V(t)$ the corresponding symplectic KvN unitary group, the previous transformations give
    \begin{equation}
        \mathfrak{U}V(t)\mathcal \mathfrak{U}^{-1}=\int_{\mathbb R}^{\oplus}U_\xi(t)\,d\xi.
    \end{equation}
    By uniqueness of the self-adjoint generators, we conclude that
    \begin{equation}
        \mathfrak{U}\overline{\hat L_h}\mathfrak{U}^{-1}=\int_{\mathbb R}^{\oplus}\overline{\hat L_{\cH,\xi}}\,d\xi,
    \end{equation}
    completing the proof.
\end{proof}
%%%%%%%%%%%%%%%%%%%%%%%%%%%%%%%%%%%%%%%%%%%%%%%%%%%%%%%%%%%%%%%%%%%%%%%%%%%%%%%%%%%%%%%%%%%%%%%%%%%%%%%%%%%%%%%%
\section{Examples}
\label{sec:KvNexamples}
%%%%%%%%%%%%%%%%%%%%%%%%%%%%%%%%%%%%%%%%%%%%%%%%%%%%%%%%%%%%%%%%%%%%%%%%%%%%%%%%%%%%%%%%%%%%%%%%%%%%%%%%%%%%%%%%
In this section we address some examples. For each of them we first introduce some context, and then we treat the Koopman von Neumann formulation from different perspectives: we use the characteristic methods to find a general solutions, and in some cases for some initial condition; while the second is an operatorial approach which will exploit Theorem~\ref{thm:diagonalization}.

%%%%%%%%%%%%%%%%%%%%%%%%%%%%%%%%%%%%%%%%%%%%%%%%%%%%%%%%%%%%%%%%%%%%%%%%%%%%%%%%%%%%%%%%%%%%%%%%%%%%%%%%%%%%%%%%
\subsection{Standard Reeb flow}
We will study the dynamics induced on $L^2(\mathbb{R}^{2n+1},\eta_{std} = ds - p_j dq^j)$ provided by the KvN operator induced by the Reeb flow, that in standard coordinates is
\begin{equation}
    \hat{K}_{-1} = -\imath \frac{\partial}{\partial s}.
\end{equation}
We can fix the integrable system by choosing $n$ Reeb-invariant functions (in literature they are called basic functions), that are linear independent. To simplify the study we can choose positions representation ($q^j$) and momenta representation (${p^j}$).
\paragraph{Momenta choice}
The complete integrable contact Hamiltonian system in this case is $(\mathbb{R}^{2n+1},\eta = ds - p_j dq^j, \{-1, p_1, \ldots p_n\})$. It is easy to verify that it both conditions in Definition~\ref{def:CCHS}. The Koopman operators involved than are:
\begin{equation}
    \hat{K}_{-1} = -\imath \frac{\partial}{\partial s} \qquad \hat{K}_{p^j} = - \imath \frac{\partial}{\partial q^j} \qquad \text{and} \qquad \hat{p^j} = p^j,
\end{equation}
with the dual basis:
\begin{equation}
    \phi_{\lambda_k, \mu_j, \kappa} = \prod_{j=1}^{n}\delta(p_j - \mu_j) e^{\imath \lambda_j q_j + \imath \kappa s}.
\end{equation}

\paragraph{Positon choice}
In similar way we might proceed with by choosing the positions, and then the contact Hamiltonian integrable system $(\mathbb{R}^{2n+1}, \eta = ds - p_j dq^j, \{-1, q_1, \ldots, q_n\})$. The integral system is easily given by:
\begin{equation}
    \hat{K}_{-1} =  -\imath \frac{\partial}{\partial s} \qquad \hat{K}_{q^j} = + \imath \frac{\partial}{\partial p^j} +\imath q^j \frac{\partial}{\partial s} \qquad \text{and} \qquad \hat{q^j} = q^j,
\end{equation}
with the dual basis:
\begin{equation}
    \phi_{\lambda_k, \mu_j, \kappa} = \prod_{j=1}^{n}\delta(q_j - \mu_j) e^{i\kappa s -\imath p_j (\lambda_j + \kappa q_j)}
\end{equation}

%%%%%%%%%%%%%%%%%%%%%%%%%%%%%%%%%%%%%%%%%%%%%%%%%%%%%%%%%%%%%%%%%%%%%%%%%%%%%%%%%%%%%%%%%%%%%%%%%%%%%%%%%%%%%%%%
\subsection{Damped Free Particle (DFP)}
In this Section, we treat the free damped particle (DFP) model. The correspondent contact Hamiltonian system is\begin{equation}\left(\mathbb{R}^{2n+1}, \eta = ds - p_i dq_i, \cH = \frac{\lvert p \rvert^2}{2} + \gamma s\right),\end{equation} 
where the equations of motion are given by:
\begin{equation}
    \begin{dcases}
        \dot{q}_i = p_i \\
        \dot{p}_i = - \gamma \, p_i \\
        \dot{s} = \frac{\lvert p \rvert^2}{2} - \gamma s
    \end{dcases}
\end{equation}
This system presents $n$ constants of motion, $f_i = \gamma q_i + p_i$, which can be derived by the first $2n$-evolution equations, and other $n$ constant of motions represented by the ratio between the $n$ momenta and the Hamiltonian
\begin{equation}
    g_k : = \frac{H}{p_k} \label{eqn:DHOconserved0}.
\end{equation}

The solution of the Hamiltonian equations are given by:
\begin{equation}
    \begin{dcases}
        p(t)_k = p_k(0) e^{-\gamma t} \\
        q(t)_k =q_k(0) + \frac{p_k(0)}{\gamma}\left( 1-e^{-\gamma t}  \right)\\
        s(t) = \frac{e^{t \gamma}-1}{2 \gamma} e^{-2 \gamma t} p_0^2 + s_0 e^{-\gamma t}  
    \end{dcases}
\end{equation}
In what follows, we restrict our attention to the contact Hamiltonian system
\begin{equation}
    \left(\mathbb{R}^3, \eta = ds - p\,dq, \cH = \frac{p^2}{2} + \gamma s \right),
\end{equation}
with $\gamma > 0$, and the contact Koopman--von Neumann formulation is given by
\begin{equation}
    \left(L^2(\mathbb{R}^3, ds - p\,dq), \hat{K}_\cH = -\imath\left(p\frac{\partial}{\partial q} - \gamma p\frac{\partial}{\partial p} + \left(\frac{p^2}{2}-\gamma s\right)\frac{\partial}{\partial s} - \gamma\mathbb{I}\right)\right).
\end{equation}
Both the Liouville equation and the Koopman--von Neumann equation can be solved using the method of characteristics, as shown in Lemma~\ref{prop:generalsolution}. Since $\reeb(\cH)=\gamma$ and $n=1$, the Liouville Cauchy problem is given by
\begin{equation}
    \begin{dcases}
        \frac{\partial \rho}{\partial t} = -p\frac{\partial \rho}{\partial q}+\gamma p\frac{\partial \rho}{\partial p}-\left(\frac{p^2}{2}-\gamma s\right)\frac{\partial \rho}{\partial s}+2\gamma\rho,\\
        \rho(0,q,p,s)=\varrho(q,p,s).
    \end{dcases}
\end{equation}
Using the characteristic curves of the contact Hamiltonian flow, its solution is
\begin{equation}
    \label{eqn:vKvNDFPcauchy}
    \rho(t,q,p,s) = e^{2\gamma t}\,\varrho\left(q+\frac{p}{\gamma}(1-e^{\gamma t}),\,pe^{\gamma t},\,se^{\gamma t}+\frac{p^2}{2\gamma}(e^{\gamma t}-e^{2\gamma t})\right).
\end{equation}

\begin{figure}[h!]
    \centering
    \includegraphics[width=0.85\textwidth]{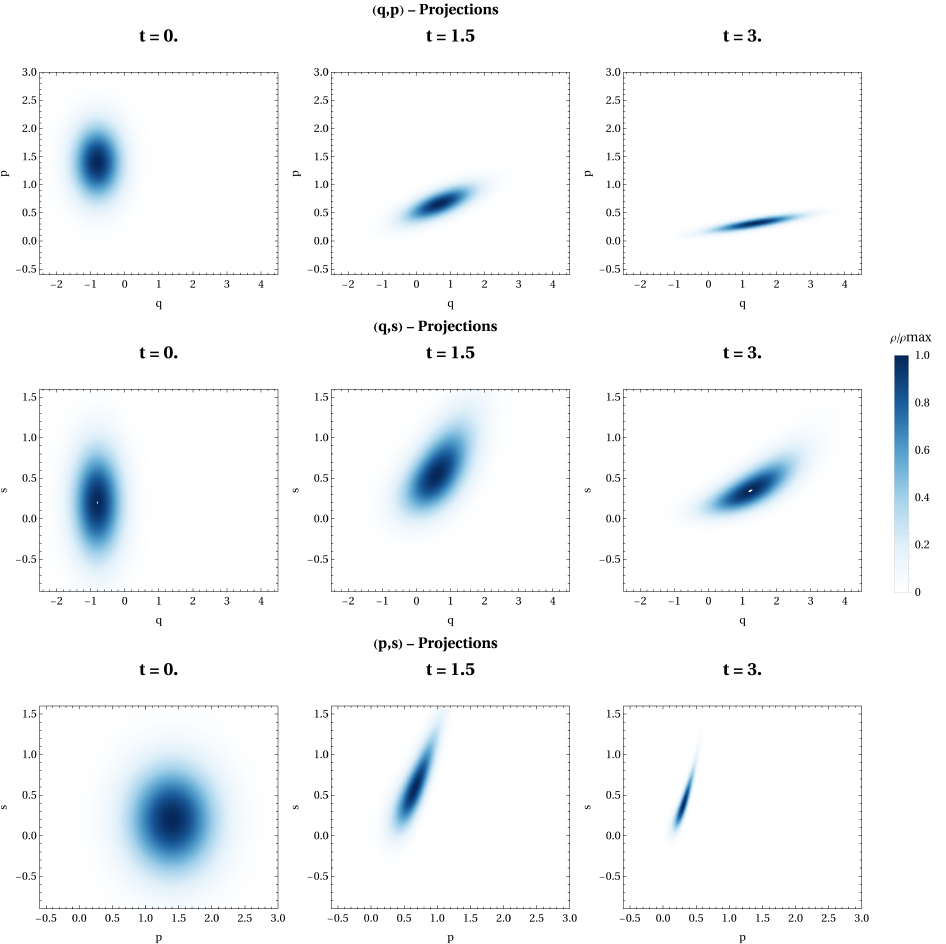}
    \caption{Time evolution of the Liouville probability density for the damped free particle, starting from a Gaussian initial condition~\eqref{eqn:initialgaussian} and $q_0 = -0.8$, $p_0 = 1.4$, $s_0 = 0.2$, $\sigma_q = 0.7$, $\sigma_p = 0.65$ and $\sigma_s = 0.65$.}
    \label{fig:DFP_evolution}
\end{figure}

Similarly, the Koopman--von Neumann Cauchy problem reads
\begin{equation}
    \begin{dcases}
        \frac{\partial \Psi}{\partial t} = -p\frac{\partial \Psi}{\partial q}+\gamma p\frac{\partial \Psi}{\partial p}-\left(\frac{p^2}{2}-\gamma s\right)\frac{\partial \Psi}{\partial s}+\gamma\Psi,\\
        \Psi(0,q,p,s)=\psi(q,p,s),
    \end{dcases}
\end{equation}
and admits the solution
\begin{equation}
    \Psi(t,q,p,s) = e^{\gamma t}\,\psi\left(q+\frac{p}{\gamma}(1-e^{\gamma t}),\,pe^{\gamma t},\,se^{\gamma t}+\frac{p^2}{2\gamma}(e^{\gamma t}-e^{2\gamma t})\right).
\end{equation}
The time evolution of the solution of the Liouville equation for the damped free particle is illustrated in Figure~\ref{fig:DFP_evolution}.

\paragraph{Gaussian initial state.}
Let us consider the normalized Gaussian initial wave function
\begin{equation}
\label{eqn:initialgaussian}
    \psi_0(q,p,s) = \frac{1}{(2\pi)^{3/4}(\sigma_q\sigma_p\sigma_s)^{1/2}}\exp\left(-\frac{(q-q_0)^2}{4\sigma_q^2}-\frac{(p-p_0)^2}{4\sigma_p^2}-\frac{(s-s_0)^2}{4\sigma_s^2}\right),
\end{equation}
where $(q_0,p_0,s_0)$ are the initial expectation values and $\sigma_q,\sigma_p,\sigma_s>0$ are the corresponding standard deviations. Since the probability density is given by $\rho_0=|\psi_0|^2$, the initial variables are independent Gaussian random variables.

The expectation values of the canonical coordinates are then given by
\begin{align}
    \langle q\rangle(t) &= q_0+\frac{1-e^{-\gamma t}}{\gamma} p_0,\\
    \langle p\rangle(t) &= e^{-\gamma t} p_0,\\
    \langle s\rangle(t) &= e^{-\gamma t} s_0+\frac{e^{-\gamma t}-e^{-2\gamma t}}{2\gamma}(p_0^2+\sigma_p^2).
\end{align}
The corresponding variances are
\begin{align}
    \sigma_q^2(t) &= \sigma_q^2+\left(\frac{1-e^{-\gamma t}}{\gamma} \right)^2\sigma_p^2,\\
    \sigma_p^2(t) &= e^{-2 \gamma t}\sigma_p^2,\\
    \sigma_s^2(t) &= e^{-2 \gamma t}\sigma_s^2+\left(\frac{e^{-\gamma t}-e^{-2\gamma t}}{2\gamma}\right)^2 (4p_0^2\sigma_p^2+2\sigma_p^4).
\end{align}

We can also compute the expectation value and variance of the contact Hamiltonian $\cH=p^2/2+\gamma s$:
\begin{equation}
    \langle\cH\rangle(t)=e^{-\gamma t}\left(\frac{p_0^2+\sigma_p^2}{2}+\gamma s_0\right),
\end{equation}
and the variance
\begin{equation}
   \sigma_\cH(t)=e^{-\gamma t}\sqrt{p_0^2\sigma_p^2+\frac{\sigma_p^4}{2}+\gamma^2\sigma_s^2}.
\end{equation}

For the Koopman--von Neumann operators, the real Gaussian state also satisfies
\begin{equation}
    \langle\hat{K}_p\rangle(t)=0,\qquad \langle\hat{K}_\cH\rangle(t)=0,\qquad \sigma_{\hat{K}_p}^2(t)=\frac{1}{4\sigma_q^2}.
\end{equation}

\begin{figure}[h!]
    \centering
    \includegraphics[width=0.85\textwidth]{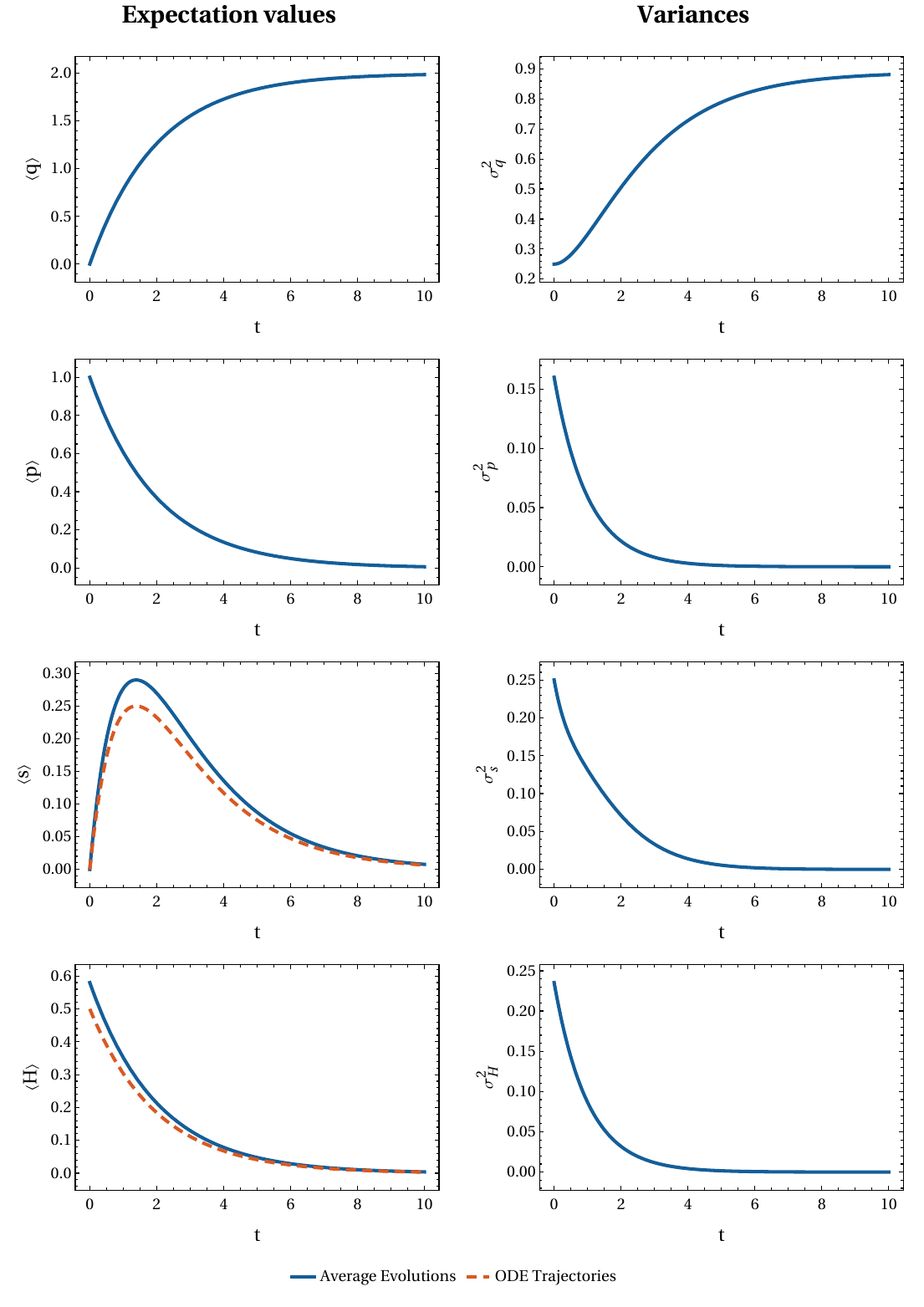}
    \caption{
        Time evolution of the expectation values (left column) and variances (right column) of $q$, $p$, $s$, and $H$ for the damped free particle. The solid blue curves represent the evolution of the averages in the Koopman formulation, while the dashed orange curves show the ODE trajectories with intial conditions given by the initial averages $q_0, p_0$ and $s_0$. The parameters are $\gamma=0.5$, $(q_0,p_0,s_0)=(0,1,0)$, and $(\sigma_q,\sigma_p,\sigma_s)=(0.5,0.4,0.5)$.
    }
    \label{fig:free_particle_moments}
\end{figure}

It is interesting to observe that the expectation values of $q$ and $p$ follow exactly the classical trajectories with initial conditions $(q_0,p_0)$. On the other hand, the expectation value of $s$ also depends on the initial momentum variance $\sigma_p^2$, reflecting the nonlinear dependence of its evolution on $p$. A similar correction appears in the expectation value of the Hamiltonian. Although the Hamiltonian is not conserved, both its expectation value and its standard deviation decay exponentially. A plot can be seen in Figure~\ref{fig:free_particle_moments}.

%%%%%%%%%%%%%%%%%%%%%%%%%%%%%%%%%%%%%%%%%%%%%%%%%%%%%%%%%%%%%%%%%%%%%%%%%%%%%%%%%%%%%%%%%%%%%%
\paragraph{Operatoria Approach.} The cKvN system associated to the DFP is
\begin{equation}
    \left(L^2(\mathbb{R}^3, ds - p\,dq), \hat{K}_\cH = -\imath\left(p\frac{\partial}{\partial q} - \gamma p\frac{\partial}{\partial p} + \left(\frac{p^2}{2}-\gamma s\right)\frac{\partial}{\partial s} - \gamma\mathbb{I}\right)\right).
\end{equation}

The diagonalization of the cKvN operator $\hat{K}_\cH$ will follow a similar fashion of~\cite{Mauro2003}, and it will exploit the complete integrability the model and Theorem~\ref{thm:diagonalization}.
A first observation comes from the fact that $\mathbb{R}^3$ is decomposed in three invariant parts: $p>0$, $p<0$ and $p=0$, this last has measure zero with respect $\Omega_M$. This permits to introduce the following set of coordinates on $\mathbb{R}^3 \setminus \{p = 0\}$:
\begin{equation}
    x=q + \frac{p}{\gamma}, \qquad y = \frac{p}{2} + \gamma \frac{s}{p}, \qquad \text{and}\quad r = \log |p|,
\end{equation}
as already shown $x$ and $y$ are constant of motion, while the evolution of $r$ is $\dot{r} = -\gamma$. With this set of coordinates the contact form, its exterior derivative and volume form become:
\begin{equation}
    \begin{dcases}
        \eta = \frac{\epsilon e^r }{\gamma} \left(dy + y dr - \gamma dx \right),\\
        d\eta = \epsilon e^r dx \wedge dr, \\
        \Omega_M = \frac{e^{2r}}{\gamma} dx dy dr
    \end{dcases}
\end{equation}
where $\epsilon = \text{sign}(p)$.
This is a contactomorphism which induces the unitary transformation:
\begin{equation}
     W:L^2(\mathbb{R}^3,\eta)\longrightarrow L^2_{+}(\mathbb{R}^3,dx\,dy\,dr) \oplus L^2_{+}(\mathbb{R}^3,dx\,dy\,dr),
\end{equation}
where $L^2_+(\mathbb{R}^3,dx\,dy\,dr)$ is associated to the positive value of the momenta, $W$ is given by:
\begin{equation}
    (W\psi)_\varepsilon(x,\kappa,r)=\frac{e^r}{\sqrt{\gamma}}\psi\left(x-\frac{\varepsilon e^r}{\gamma},\,\varepsilon e^r,\,\frac{\varepsilon e^r}{\gamma}\left(\kappa-\frac{\varepsilon e^r}{2}\right)\right).
\end{equation}
Now the operators $\hat{K}_\cH$, $\hat{K}_p$, $\hat{f}$ become
\begin{equation}
    \label{eqn:transformedoperatorsDFP}
    W\hat{K}_\cH W^{-1}=\imath\gamma\partial_r,\qquad W\hat{K}_pW^{-1}=-\imath\partial_x,\qquad W\hat{f}W^{-1}=y.
\end{equation}
\begin{remark}
    The diagonalization of $\hat{K}_\cH$ and $\hat{K}_p$ reduces respectively to the Fourier transformation in the variables $r$ and $x$.
\end{remark}

\begin{Prop}
    \label{prop:DFPspectrum}
    The cKvN operator $\hat{K}_\cH$ of the DFP admits a complete generalized spectral decomposition on $L^2(\mathbb{R}^3,\eta)$. The joint generalized eigenfunctions of $\hat{K}_\cH$, $\hat{K}_p$, $\hat{f}$ are given, on $p\neq0$, by
    \begin{equation}
    \phi_{\mu,\lambda,\kappa}^{\pm}(q,p,s) = \frac{\mathbf{1}_{\{\pm p>0\}}}{2\pi|p|} \exp\left(\imath\lambda\left(q+\frac{p}{\gamma}\right)-\imath\frac{\mu}{\gamma}\log|p|\right) \delta\left(\frac{p}{2}+\gamma\frac{s}{p}-\kappa\right).
    \end{equation}
    where $\mu,\lambda,\kappa\in\mathbb{R}$. The corresponding eigenvalue equations are
    \begin{equation}
        \hat{K}_\cH\phi_{\mu,\lambda,\kappa,\varepsilon}=\mu\phi_{\mu,\lambda,\kappa,\varepsilon},\qquad \hat{K}_p\phi_{\mu,\lambda,\kappa,\varepsilon}=\lambda\phi_{\mu,\lambda,\kappa,\varepsilon}, \qquad \hat{f}\phi_{\mu,\lambda,\kappa,\varepsilon}=\kappa\phi_{\mu,\lambda,\kappa,\varepsilon}.
    \end{equation}
    These generalized eigenfunctions satisfy the distributional orthogonality relation
    \begin{equation}
        \langle\phi_{\mu,\lambda,\kappa,\varepsilon},\phi_{\mu',\lambda',\kappa',\varepsilon'}\rangle=\delta(\mu-\mu')\delta(\lambda-\lambda')\delta(\kappa-\kappa'),
    \end{equation}
    and provide a resolution of the identity on $L^2(\mathbb{R}^3,\eta)$.
\end{Prop}
\begin{proof}
    The operators in equation~\eqref{eqn:transformedoperatorsDFP} commutes since the original operators commute for Theorem~\ref{thm:KvNchar} point~\ref{thm:KvncharLie}. Then by using the Fourier transformation we can find the generalized eigenfunctions of the transformed operators are:
    \begin{equation}
        \Phi_{\mu,\lambda,\kappa}(x,y,r)=\frac{1}{2\pi\sqrt{\gamma}}e^{\imath\lambda x-\imath\frac{\mu}{\gamma}r}\delta(y-\kappa).
    \end{equation}
    The eigen relations can be easily checked by direct computation, while the normalization follows from the Fourier identities
    \begin{equation}
        \int_{\mathbb{R}}e^{\imath(\lambda-\lambda')x}\,dx=2\pi\delta(\lambda-\lambda'),\qquad \int_{\mathbb{R}}e^{\imath\frac{\mu-\mu'}{\gamma}r}\,dr=2\pi\gamma\delta(\mu-\mu'),
    \end{equation}
    to obtain
    \begin{equation}
        \langle\Phi_{\mu,\lambda,\kappa},\Phi_{\mu',\lambda',\kappa'}\rangle=\delta(\mu-\mu')\delta(\lambda-\lambda')\delta(\kappa-\kappa').
    \end{equation}
    Now applying the unitary transformation $W^{-1}$ to the generalized eigenfunctions and restricting to one of the two invariant set of the momentum:
    \begin{equation}
    \phi_{\mu,\lambda,\kappa}^{\pm}(q,p,s) = \frac{\mathbf{1}_{\{\pm p>0\}}}{2\pi|p|} \exp\left(\imath\lambda\left(q+\frac{p}{\gamma}\right)-\imath\frac{\mu}{\gamma}\log|p|\right) \delta\left(\frac{p}{2}+\gamma\frac{s}{p}-\kappa\right),
    \end{equation}
    and since the transformation $W$ is unitary normalization and the resolution of the identity is preserved.
\end{proof}
The spectral decomposition permits to express any initial KvN wave function $\psi \in L^2(\mathbb{R}^3, \eta)$ in terms of a set of complex coefficients:
\begin{equation}
    c^{\pm}[\psi]_{\mu,\lambda,\kappa}=\int_{\mathbb{R}^3}\overline{\phi_{\mu,\lambda,\kappa}^{\pm}(q,p,s)}\,\psi(q,p,s)\,dq\,dp\,ds, 
\end{equation}
and use them to compute the time evolution:
\begin{align}
    \Psi(t,q,p,s)=& \int_{\mathbb{R}^3}e^{-\imath\mu t}\, c^+[\psi]_{\mu,\lambda,\kappa} \, \phi_{\mu,\lambda,\kappa}^{+}(q,p,s)\,d\mu\,d\lambda\,d\kappa \\ 
    & \qquad + \int_{\mathbb{R}^3}e^{-\imath\mu t}\, c^-[\psi]_{\mu,\lambda,\kappa}\, \phi_{\mu,\lambda,\kappa}^{-}(q,p,s)\,d\mu\,d\lambda\,d\kappa.
\end{align}
\begin{remark}
    Let us consider the Cauchy problem~\eqref{eqn:vKvNDFPcauchy} with the Gaussian initial condition~\eqref{eqn:initialgaussian}. The spectral coefficients can be computed explicitly with respect to $q$ and $s$, leaving a one-dimensional integral:
    \begin{align}
        c_{\mu,\lambda,\kappa}^{\pm}
        &=\frac{\sigma_q e^{-\sigma_q^2\lambda^2-\imath\lambda q_0}}{\sqrt{\pi}\gamma(2\pi)^{3/4}(\sigma_q\sigma_p\sigma_s)^{1/2}}
        \int_0^\infty u^{\imath\mu/\gamma}
        \exp\left[-\frac{(\pm u-p_0)^2}{4\sigma_p^2}-\frac{\left(\pm\kappa u/\gamma-u^2/(2\gamma)-s_0\right)^2}{4\sigma_s^2}-\imath\pm\frac{\lambda u}{\gamma}\right]du,
    \end{align}
    where the remaining integral can be evaluated numerically. Despite the positive and negative momentum spaces are invariant and independent, the Gaussian initial condition has nonzero support in both. Their respective probabilities are
    \begin{equation}
        \mathbb{P}(\text{sign}(p)= \pm 1) = \frac12\left(1+\pm \operatorname{erf}\left(\frac{p_0}{\sqrt{2}\sigma_p}\right)\right).
    \end{equation}
    and they are conserved by the time evolution generated by $\hat{K}_\cH$.
\end{remark}

%%%%%%%%%%%%%%%%%%%%%%%%%%%%%%%%%%%%%%%%%%%%%%%%%%%%%%%%%%%%%%%%%%%%%%%%%%%%%%%%%%%%%%%%%%%%%%
\subsection{Damped Harmonic Oscillator}
\label{subsec:DHO}

The damped harmonic oscillator is another classical example of a dissipative system that admits a contact Hamiltonian formulation~\cite{Bravetti2017}. We consider the system
\begin{equation}
    \left(\mathbb{R}^3,\eta=ds-pdq,\cH=\frac{p^2}{2}+\frac{q^2}{2}+\gamma s\right),
\end{equation}
where $\gamma>0$. The corresponding contact Hamiltonian vector field and cKvN operator are
\begin{equation}
    X_\cH=p\partial_q-(q+\gamma p)\partial_p+
    \left(\frac{p^2}{2}-\frac{q^2}{2}-\gamma s\right)\partial_s,
    \qquad
    \hat K_\cH=-\imath X_\cH+\imath\gamma.
\end{equation}
The equations of motion are
\begin{equation}
    \dot q=p,\qquad
    \dot p=-q-\gamma p,\qquad
    \dot s=\frac{p^2}{2}-\frac{q^2}{2}-\gamma s.
\end{equation}
Since the first two equations form a linear system and the third is a linear inhomogeneous equation in $s$, the Hamiltonian vector field is complete. Moreover, the function
\begin{equation}
    f=2s-qp
\end{equation}
is in Jacobi involution with $\cH$, and its Hamiltonian vector field is
\begin{equation}
    X_f=-q\partial_q-p\partial_p-2s\partial_s.
\end{equation}
This vector field generates an anisotropic scaling centred at the origin of $\mathbb R^3$ and is also complete. The independence condition of Definition~\ref{def:CCHS} holds on an open subset of full measure. Thus, the system is completely integrable in the contact sense.

To obtain an explicit spectral decomposition, we restrict to the underdamped regime $0<\gamma<2$ and introduce a change of coordinates $\chi:(u,\theta,y)\mapsto(q,p,s)$ defined by
\begin{equation}
    \begin{aligned}
        \omega&=\sqrt{1-\frac{\gamma^2}{4}},\\
        q&=e^u\cos\theta,\\
        p&=e^u\left(\omega\sin\theta-\frac{\gamma}{2}\cos\theta\right),\\
        s&=e^{2u}\left(y+\frac{\omega}{2}\sin\theta\cos\theta-\frac{\gamma}{4}\cos^2\theta\right).
    \end{aligned}
\end{equation}
This transformation defines a diffeomorphism between $\mathbb R\times S^1\times\mathbb R$ and $\mathbb R^3\setminus\{q=p=0\}$. In these coordinates, the Hamiltonian vector fields become
\begin{equation}
    X_\cH=-\frac{\gamma}{2}\partial_u-\omega\partial_\theta,
    \qquad
    X_f=-\partial_u.
\end{equation}
Moreover,
\begin{equation}
    \cH=e^{2u}\left(\frac{\omega^2}{2}+\gamma y\right),
    \qquad
    f=2e^{2u}y,
    \qquad
    \cH\,df-f\,d\cH=\omega^2e^{4u}dy.
\end{equation}
Consequently, the regular set is $M_{\mathrm{reg}}=\mathbb R^3\setminus\{q=p=0\}$, and $y$ is conserved by both Hamiltonian flows.

The contact volume form becomes
\begin{equation}
    \Omega_M=\omega e^{4u}\,du\wedge d\theta\wedge dy.
\end{equation}
Therefore, the transformation
\begin{equation}
    (W\psi)(u,\theta,y)
    =\sqrt{\omega}\,e^{2u}(\psi\circ\chi)(u,\theta,y)
\end{equation}
defines a unitary map
\begin{equation}
    W:L^2(M,\Omega_M)
    \longrightarrow
    L^2(\mathbb R\times S^1\times\mathbb R,du\,d\theta\,dy).
\end{equation}
The corresponding cKvN operators become
\begin{equation}
    W\hat K_\cH W^{-1}
    =\imath\left(\frac{\gamma}{2}\partial_u+\omega\partial_\theta\right),
    \qquad
    W\hat K_f W^{-1}=\imath\partial_u,
    \qquad
    W\hat y W^{-1}=y,
\end{equation}
where $\hat y$ denotes multiplication by the conserved function $y$.

These operators admit the joint generalized eigenfunctions
\begin{equation}
    \widetilde{\phi}_{\lambda,m,\kappa}(u,\theta,y)
    =\frac{1}{2\pi}e^{-\imath\lambda u+\imath m\theta}\delta(y-\kappa),
\end{equation}
where $\lambda,\kappa\in\mathbb R$ and $m\in\mathbb Z$. The corresponding generalized eigenfunctions in the original contact Hilbert space, expressed in the adapted coordinates, are
\begin{equation}
    (\phi_{\lambda,m,\kappa}\circ\chi)(u,\theta,y)
    =\frac{1}{2\pi\sqrt{\omega}}e^{-2u}
    e^{-\imath\lambda u+\imath m\theta}\delta(y-\kappa).
\end{equation}
Returning to the original coordinates $(q,p,s)$, we obtain
\begin{equation}
    \begin{aligned}
    \phi_{\lambda,m,\kappa}(q,p,s)
    ={}&\frac{1}{2\pi\sqrt{\omega}}
    \exp\left[
        -\frac{\imath\lambda}{2}
        \log\left(q^2+\frac{(p+\gamma q/2)^2}{\omega^2}\right)
        +\imath m\arg\left(q+\imath\frac{p+\gamma q/2}{\omega}\right)
    \right]\\
    &\times
    \delta\left(
        s-\frac{qp}{2}
        -\kappa\left(q^2+\frac{(p+\gamma q/2)^2}{\omega^2}\right)
    \right).
    \end{aligned}
\end{equation}
These generalized eigenfunctions satisfy
\begin{equation}
    \hat K_\cH\phi_{\lambda,m,\kappa}
    =\left(\frac{\gamma\lambda}{2}-\omega m\right)\phi_{\lambda,m,\kappa},
    \qquad
    \hat K_f\phi_{\lambda,m,\kappa}
    =\lambda\phi_{\lambda,m,\kappa},
    \qquad
    \hat y\phi_{\lambda,m,\kappa}
    =\kappa\phi_{\lambda,m,\kappa}.
\end{equation}
%%%%%%%%%%%%%%%%%%%%%%%%%%%%%%%%%%%%%%%%%%%%%%%%%%%%%%%%%%%%%%%%%%%%%%%%%%%%%%%%%%%%%%%%%%%%%%
\section{Conclusions}
In this work, we have developed a Koopman-von Neumann formulation of contact Hamiltonian systems. While extensions of KvN mechanics to non-Hamiltonian dynamics are already known~\cite{Joseph2020}, our approach focuses on the geometric structures underlying contact dynamics. We have characterized the cKvN generator and its unitary evolution, established its relation with contact integrability and spectral decomposition, and investigated the evolution of differential entropy. Furthermore, we have related the cKvN formulation to the standard symplectic KvN framework through symplectization and illustrated the results with several examples.

Form this formulation, several directions for future research arise: one possibility is to extend the approach to more general geometric structures, such as Poisson and Jacobi manifolds or sub-Riemannian geometries, where the choice of a suitable volume form plays an important role. Another direction concerns the composition and coupling of contact Hamiltonian systems. In particular, the tensor product of cKvN Hilbert spaces provides a natural construction for independent systems. This could also motivate the development of structure-preserving numerical methods for coupled dissipative systems. Finally, the present Hilbert-space formulation offers a possible starting point for studying contact quantization~\cite{Gozzi1988,Gozzi1989} and classical-quantum hybrid dynamics, building on existing approaches to contact geometric quantization and Koopman--van Hove mechanics~\cite{Fitzpatrick2011,GayBalmazTronci2022}.

%%%%%%%%%%%%%%%%%%%%%%%%%%%%%%%%%%%%%%%%%%%%%%%%%%%%%%%%%%%%%%%%%%%%%%%%%%%%%%%%%%%%%%%%%%%%%%

\paragraph{Acknowledgments.} The research of the author was partially supported by FWO-EoS project Beyond symplectic geometry with UA Antigoon number 45816 and the grant Francqui Research Professor 2023-2026 of the Francqui Foundation with UA Antigoon number 49741. 

%%%%%%%%%%%%%%%%%%%%%%%%%%%%%%%%%%%%%%%%%%%%%%%%%%%%%%%%%%%%%%%%%%%%%%%%%%%%%%%%%%%%%%%%%%%%%%
\printbibliography

\appendix
%%%%%%%%%%%%%%%%%%%%%%%%%%%%%%%%%%%%%%%%%%%%%%%%%%%%%%%%%%%%%%%%%%%%%%%%%%%%%%%%%%%%%%%%%%%%%
\section{Symplectization}
\label{app:symplectization}

Symplectization associates a symplectic manifold to a contact manifold, allowing contact Hamiltonian dynamics to be described within a symplectic framework~\cite{Arnold_1989}.

Given a contact manifold $(M,\Delta)$, its intrinsic symplectization is the nonzero annihilator of the contact distribution,
\begin{equation}\label{eq:intrinsic_sm}
    SM=\left\{(x,\lambda)\in T^*M:\lambda\neq0,\quad \lambda|_{\Delta_x}=0\right\},
\end{equation}
equipped with the restriction of the canonical symplectic form $\omega_{T^*M}=-d\theta$. If the contact structure is coorientable, both components of $SM$ can be identified with $\mathbb{R}\times M$. The standard construction can be obtained by the identification $(r,x)\mapsto(x,-e^r\eta_x)$.

\begin{theorem}\label{thm:symplectization_global}
    Let $(M,\eta)$ be a cooriented contact manifold of dimension $2n+1$. Then $\mathbb{R}\times M$, equipped with
    \begin{equation}\label{eq:global_omega}
        \omega=d(e^r\eta)=e^r(dr\wedge\eta+d\eta),
    \end{equation}
    is an exact symplectic manifold of dimension $2n+2$. Moreover, $W=\partial_r$ is a Liouville vector field satisfying $\Lie_W\omega=\omega$.
\end{theorem}

\begin{proof}
    Since $\omega$ is exact, it is closed. Its non-degeneracy follows from the volume form
    \begin{equation}
        \label{eqn:sympvolume}
        \Omega_SM = \omega^{n+1}=e^{(n+1)r}dr\wedge\eta\wedge(d\eta)^n\neq0,
    \end{equation}
    by the contact condition. Furthermore, $\iota_W\omega=e^r\eta$, and Cartan's formula gives
    \begin{equation}
        \Lie_W\omega=d(\iota_W\omega)=d(e^r\eta)=\omega.
    \end{equation}
\end{proof}

The following proposition characterizes the symplectic Hamiltonians whose vector fields are invariant under the Liouville flow.

\begin{Prop}\label{prop:hamiltonian_homogeneity}
    Let $h\in C^\infty(\mathbb{R}\times M)$, its symplectic Hamiltonian vector field $V_h$ is defined by equation~\eqref{eqn:sympHVF}. Assuming $M$ is connected, $[W,Y_h]=0$ if and only if
    \begin{equation}\label{eq:global_homogeneity}
        h(r,x)=c_1+e^r f(x),
    \end{equation}
    where $c_1\in\mathbb{R}$ and $f\in C^\infty(M)$. Moreover, the Hamiltonian vector field of $h=e^r f$ is projected onto the contact Hamiltonian vector field $X_f$ on $M$.
\end{Prop}

\begin{proof}
    Taking the Lie derivative of $\omega(Y_h, \cdot)=dh$ along $W$ and using $\Lie_W\omega=\omega$, we obtain
    \begin{align}
        \omega([W, Y_h], \cdot)&=\mathcal{L}_W \omega(Y_h, \cdot) - (\mathcal{L}_W \omega)(Y_h, \cdot) \\
        &=\mathcal{L}_W dh - \omega(Y_h,\cdot) \\
        &=d(W(h) - h)
    \end{align}
    The commutator vanishes, i.e. $[W,Y_h]=0$, if and only if $W(h)-h$ is constant, and the integration by parts gives $h=c_1+e^r \cH$.

    For $h=e^r f$, using the equations~\eqref{eqn:defHVF}, and $\omega$ in the form~\eqref{eq:global_omega}, it is possible to show that
    \begin{equation}
        \omega(X_\cH+\reeb(\cH)\partial_r, \cdot)=d(e^r \cH),
    \end{equation}
    that means: $Y_h=X_\cH+\reeb(\cH)\partial_r$, and $\pi_*Y_h=X_\cH$, where $\pi:\mathbb{R}\times M\to M$ is the canonical projection.
\end{proof}

\paragraph{Canonical Expression of symplectization.} Consider a contact Hamiltonian system in Darboux coordinates, with $\eta=ds-p_i\,dq^i$ and the Hamiltonian function $\cH(q^i,p_i,s)$. Its symplectization gives the symplectic Hamiltonian system:
\begin{equation}
    \left(\mathbb{R}^{2n+2}, \omega=e^r\left(dr\wedge ds-p_i\,dr\wedge dq^i+dq^i\wedge dp_i\right), h=e^r \cH\right)
\end{equation}
The associated symplectic Hamiltonian vector field is
\begin{equation}
    Y_h=\frac{\partial \ch}{\partial p_i}\partial_{q^i}
    -\left(\frac{\partial\cH}{\partial q^i}+p_i\frac{\partial\cH}{\partial s}\right)\partial_{p_i}
    +\left(p_i\frac{\partial\cH}{\partial p_i}-f\right)\partial_s
    +\frac{\partial\cH}{\partial s}\partial_r.
\end{equation}
Its projection onto $M$ recovers the contact Hamiltonian vector field $X_\cH$.

Although $\omega$ is not in symplectic Darboux form in these coordinates, it can be written in standard symplectic coordinates. In dimension four, the transformation
\begin{equation}
    Q_1=r,\qquad P_1=e^r(s-qp),\qquad Q_2=qe^r,\qquad P_2=p
\end{equation}
with
\begin{equation}
    \omega=dQ_1\wedge dP_1+dQ_2\wedge dP_2.
\end{equation}
The corresponding symplectic Hamiltonian becomes
\begin{equation}
    h(Q_1,Q_2,P_1,P_2)=e^{Q_1}\cH\left(Q_2e^{-Q_1},P_2,e^{-Q_1}(P_1+P_2Q_2)\right).
\end{equation}
\end{document}